\documentclass[twocolumn,preprintnumbers,superscriptaddress,amsmath,amssymb]{revtex4}
\usepackage{float}
\usepackage{graphicx}
\usepackage{dcolumn}
\usepackage{bm}
\usepackage{epstopdf}
\usepackage{xcolor}
\usepackage{textcmds}
\usepackage{textcomp}
\usepackage{soul}
\usepackage{csquotes}
\usepackage{amsbsy}
\usepackage{mathrsfs} 
\usepackage{amsmath}
\usepackage{bigints} 
\usepackage{amsthm,amssymb} 
\usepackage[utf8]{inputenc}
\usepackage[english]{babel}
\usepackage[shortlabels]{enumitem}
\usepackage{float}
\usepackage{mathrsfs,bigints,mathtools,dsfont}
\usepackage[colorlinks=true,linkcolor=blue,citecolor=blue]{hyperref}%
\usepackage[toc]{appendix}
\newcommand\norm[1]{\left\lVert#1\right\rVert}

\begin{document}
	
 \title{{ Collective dynamics of swarmalators with higher-order interactions  }}
 
 \author{Md Sayeed Anwar}\affiliation{Physics and Applied Mathematics Unit,
Indian Statistical Institute, 203 B. T. Road, Kolkata 700108, India}
\author{Gourab Kumar Sar}\affiliation{Physics and
Applied Mathematics Unit, Indian Statistical Institute, 203 B. T. Road, Kolkata 700108, India}
\author{Matja{\v z} Perc}\affiliation{Faculty of Natural Sciences and Mathematics, University of Maribor,
Koro{\v s}ka cesta 160, 2000 Maribor, Slovenia}\affiliation{Department of Medical Research, China Medical University Hospital, China Medical University, Taichung 404332, Taiwan}\affiliation{ Alma Mater Europaea, Slovenska ulica 17, 2000 Maribor, Slovenia}\affiliation{Complexity Science Hub Vienna, Josefst{\"a}dterstra{\ss}e 39, 1080 Vienna, Austria}\affiliation{Department of Physics, Kyung Hee University, 26 Kyungheedae-ro, Dongdaemun-gu, Seoul, Republic of Korea}
\author{Dibakar Ghosh}\email{dibakar@isical.ac.in}\affiliation{Physics and
Applied Mathematics Unit, Indian Statistical Institute, 203 B. T. Road, Kolkata 700108, India} 	
    
\begin{abstract}
Higher-order interactions shape collective dynamics, but how they affect transitions between different states in swarmalator systems is yet to be determined. To that effect, we here study an analytically tractable swarmalator model that incorporates both pairwise and higher-order interactions, resulting in four distinct collective states: async, phase wave, mixed, and sync states.  We show that even a minute fraction of higher-order interactions induces abrupt transitions from the async state to the phase wave and the sync state. We also show that higher-order interactions facilitate an abrupt transition from the phase wave to the sync state by bypassing the intermediate mixed state. Moreover, elevated levels of higher-order interactions can sustain the presence of phase wave and sync state, even when pairwise interactions lean towards repulsion. The insights gained from these findings unveil self-organizing processes that hold the potential to explain sudden transitions between various collective states in numerous real-world systems.   
\end{abstract}
	\maketitle	
\section{Introduction} \label{intro}
The dual interplay between swarming and synchronization is in the heart of \textit{swarmalation} phenomena which is commonly used to delineate the collective behaviors of entities called \textit{swarmalators}. With ever-growing advancements and discoveries in multi-agent system studies, it has been observed that there are numerous systems where entities aggregate in space and synchronize over time. Such systems, both natural and man-made, are appropriate examples of swarmalator systems that exhibit simultaneous swarming and synchronization effects. Japanese tree frogs~\cite{ota2020interaction}, magnetic domain walls~\cite{hrabec2018velocity}, swarming robots~\cite{barcis2020sandsbots}, magnetotactic bacteria~\cite{belovs2017synchronized}, vinegar eels~\cite{peshkov2022synchronized}, Quincke rollers~\cite{zhang2020reconfigurable}, Janus particles~\cite{yan2012linking} are few examples where swarmalation effects are encountered.
\par Although the two fields synchronization~\cite{pikovsky2001universal,boccaletti2001unifying} and swarming~\cite{sumpter2010collective,bialek2012statistical,reynolds1987flocks} have been extensively explored over the last few decades, studies on their combined effect began not very long ago. In the seminal work of Vicsek et. al~\cite{vicsek1995novel}, particles moved inside a bounded region, and their directions were influenced by the neighboring particles lying inside a unit radius. Despite being novel to illustrate the phase transition from a desynchronized to a synchronized state, the Vicsek model places little focus on the spatial position and structures of the particles. Later, depending on the spatial movement of the particles, synchronization phenomena were studied by the introduction of mobile agents or moving oscillators~\cite{stilwell2006sufficient,frasca2008synchronization,uriu2013dynamics}. Here also, the effect of spatial position and internal dynamics is unidirectional: the position of the particles influences their internal dynamics, but not the other way round. The pivotal works that laid the platform for the swarmalator systems were carried out by Tanaka et al.~\cite{tanaka2007general} and Isawa et al.~\cite{iwasa2010dimensionality,iwasa2010hierarchical} while studying the movement and dynamics of chemotactic oscillators. The movements of these oscillators are mediated by the surrounding chemical. In 2017, O'Keeffe et al.~\cite{o2017oscillators} proposed a simple mathematical model of swarmalators where they move in the two-dimensional plane with Kuramoto-like oscillator dynamics. Five long-term collective states for position aggregation and phase synchronization, viz., {\it static sync}, {\it static phase wave}, {\it splintered phase wave}, {\it active phase wave}, and {\it static async} were reported in this study. The spatial attraction between two swarmalators was affected by their relative phase, and the spatial distance between them influenced the phase coupling. Adopting this central idea of mutual influence of the spatial position and phase, this model has further been studied with different interaction functions~\cite{o2018ring,jimenez2020oscillatory,ceron2023diverse,hong2018active}, coupling schemes~\cite{hong2021coupling,sar2022swarmalators,lee2021collective}, external forcing~\cite{lizarraga2020synchronization}, large particle limit~\cite{ha2021mean,ha2019emergent}, etc.~\cite{o2019review,sar2022dynamics} A plethora of new collective states has been found which in turn made the researchers interested in delving deeper into the study of such systems. However, mathematical analysis in terms of the solvability of the models or analytical properties of the emerging states was lacking in most of the cases.
\par To address this problem, O'Keeffe et al.~\cite{o2022collective,yoon2022sync} came up with a 1D swarmalator model which looked like a pair of coupled Kurmaoto equations,
\begin{subequations}\label{eq1}    
	\begin{equation} \label{1a}
		\begin{array}{cc}
			\dot{x}_{i}= v_{i}+\dfrac{J}{N} \sum\limits_{j=1}^{N} \sin(x_{j}-x_{i}) \cos(\theta_{j}-\theta_{i}) ,
		\end{array}
	\end{equation}
	\begin{equation}  \label{1b}
		\begin{array}{cc}
			\dot{\theta}_{i}= \omega_{i}+\dfrac{K}{N} \sum\limits_{j=1}^{N} \sin(\theta_{j}-\theta_{i}) \cos(x_{j}-x_{i}) , 
		\end{array}
	\end{equation}
\end{subequations}
where $x_i$ and $\theta_i$ represent the position on a 1D ring and the phase of the $i^{{th}}$ swarmalator, respectively for $i=1,2,\ldots,N$. $v_i$, $\omega_i$ are the velocity and internal frequency, and $J,K$ are the inter-element coupling strengths. Being similar to the Kuramoto model, this model was analytically tractable. Analyses of the emerging states had been possible when the model is studied with nonidentical velocities and frequencies~\cite{yoon2022sync}, distributed couplings~\cite{o2022swarmalators}, random pinning~\cite{sar2023pinning,sar2023solvable}, thermal noise~\cite{hong2023swarmalators}, phase lags~\cite{lizarraga2023synchronization}, etc.
\par All these prior research on swarmalators has predominantly focused on a thorough exploration of their behavior within the framework of pairwise interactions among the constituent entities of the system. More precisely, the spatial and phase dynamics that dictate the interactions among swarmalators have been exclusively governed by the presence of pairwise connections that link them together. Nevertheless, the reliance on a hypothesis rooted solely in pairwise interactions proves inadequate in capturing a wide array of pertinent scenarios \cite{ludington2022higher,swain2022higher,garaud2015vortex}. An example includes the group interactions among microorganisms in microbial communities \cite{ludington2022higher}. This limitation becomes especially apparent in cases where the interplay of phase and spatial dynamics among individual swarmalators is not solely influenced by pairwise connections among swarmalators but rather by the simultaneous influence of multiple interconnected swarmalators. This intricate interdependence cannot be adequately decomposed solely into the framework of pairwise connections and thus necessitates the introduction of higher-order (group) interactions \cite{ludington2022higher,battiston2021physics,boccaletti2023structure,battiston2020networks,majhi2022dynamics}. 
\par Recent advances in physics and other communities have drawn specific attention to the significance of interactions among dynamic units that extend beyond the pairwise realm. Notably, three- and four-way interactions have come to the forefront, revealing their pivotal role in shaping collective behaviors \cite{zhang2023higher,skardal2020higher,alvarez2021evolutionary}. As a result, the field of network science has shifted its focus toward comprehending higher-order structures to more accurately capture the diverse interactions that exist beyond conventional pairwise connections \cite{battiston2021physics,boccaletti2023structure,battiston2020networks,majhi2022dynamics}. These intricate interactions are frequently encoded within simplicial complexes \cite{bianconi2021higher,bick2023higher,giusti2016two}, delineating various levels of simplex structures within the network. An assemblage of 1-simplices (edges/links), 2-simplices (filled triangles), and so on constitute the intricate framework of the simplicial complex, reflecting the essence of these higher-order interactions.
\par In this context, the impact of higher-order interactions on the domain of synchronization has been the subject of thorough investigation in recent years \cite{kovalenko2021contrarians,kachhvah2022hebbian,ghorbanchian2021higher,anwar2022intralayer,lucas2020multiorder,
anwar2022stability,gambuzza2021stability,anwar2023synchronization,gallo2022synchronization,anwar2023neuronal}. These studies have unveiled that the incorporation of higher-order interactions among dynamic units has the potential to give rise to a plethora of new collective phenomena. However, the influence of higher-order interactions on the realm of swarmalators, which is indissolubly linked to the field of synchronization, remains an unexplored territory to date, and therefore, it is imperative to investigate this uncharted territory.   
\par Motivated by this, in this paper, we propose a model of swarmalators that encompasses both pairwise and higher-order interactions, notably three-body interactions among the swarmalators. These interactions are intricately woven into a simplicial complex framework at the microscopic level. Our proposed model extends the 1D swarmalator model [Eq.~\eqref{eq1}] on a ring to the framework that incorporates higher-order interactions among the phase and space dynamics of the swarmalators and thus analytically tractable using the generalized Ott-Antonsen (OA) ansatz \cite{ott2008low} in the thermodynamic $(N \to \infty)$ limit. Similar to the pairwise model, the present model also displays a diverse range of dynamics, featuring four distinct collective states: async, phase wave, mixed, and sync states. We aim to understand how higher-order interactions influence the formation and characteristics of these distinct collective states. Due to the introduction of higher-order interactions, several significant phenomena arise that are absent when swarmalator interactions are limited to only pairwise connections. Our observations highlight that the inclusion of higher-order interactions leads to abrupt transitions from the async state to the phase wave state and the sync state, contingent on the specific configurations of coupling strengths. We also observe that stronger higher-order interactions can give rise to the persistence of phase wave and sync states, even in cases where pairwise couplings are negative (i.e., repulsive). Furthermore, our findings also reveal that substantial higher-order couplings can facilitate a direct emergence of the synchronized state from the phase wave state, bypassing the intermediate mixed state.
\section{Model} \label{model}
We consider an ensemble of $N$ swarmalators subjected to two- and three-body interactions, embedded in a simplicial complex at the microscopic level, where the instantaneous position and phase of the $i^{th}$ swarmalator are represented by $(x_{i},\theta_{i}) \in (\mathbb{S}^{1},\mathbb{S}^{1})$. When decoupled, each swarmalator is characterized by a set of natural velocity and frequency $(v_{i},\omega_{i})$, drawn from a specific distribution $g_{v,\omega}$. For the sake of pedagogy, we here choose the intrinsic frequencies to be drawn from a Lorentzian distribution, $g_{v,\omega}(x)=\frac{\Delta_{v,\omega}}{\pi(x^{2}+\Delta_{v,\omega}^{2})}$, with zero mean and half-width $\Delta_{v,\omega}$. The evolution of the swarmalators under the impression of pairwise and triadic interactions is then given by, 
\begin{widetext}
\begin{subequations}    
\begin{equation} \label{sw_position_eq}
\begin{array}{cc}
    \dot{x}_{i}= v_{i}+\dfrac{J_1}{N} \sum\limits_{j=1}^{N} \sin(x_{j}-x_{i}) \cos(\theta_{j}-\theta_{i}) +\dfrac{J_2}{N^2} \sum\limits_{j=1}^{N} \sum\limits_{k=1}^{N} \sin(2x_{j}-x_{k}-x_{i}) \cos(2\theta_{j}-\theta_{k}-\theta_{i}),
\end{array}
\end{equation}
\begin{equation}  \label{sw_phase_eq}
\begin{array}{cc}
    \dot{\theta}_{i}= \omega_{i}+\dfrac{K_1}{N} \sum\limits_{j=1}^{N} \sin(\theta_{j}-\theta_{i}) \cos(x_{j}-x_{i}) +\dfrac{K_2}{N^2} \sum\limits_{j=1}^{N} \sum\limits_{k=1}^{N} \sin(2\theta_{j}-\theta_{k}-\theta_{i}) \cos(2x_{j}-x_{k}-x_{i}), 
\end{array}
\end{equation}
\end{subequations}
\end{widetext}
where $(J_{1}, K_{1})$ and $(J_{2}, K_{2})$ are the pairwise and triadic coupling strengths associated with the spatial and phase interactions, respectively. Note that when $(J_{2}, K_{2})=(0,0)$, it coincides with the conventional evolution equation of swarmalators over a ring \cite{yoon2022sync}. Therefore, analogous to the pairwise swarmalators model over a ring, Eq. \eqref{sw_phase_eq} incorporates position-dependent synchronization in swarmalators. To achieve synchronization, we employ the well-known Kuramoto sine terms \cite{acebron2005kuramoto,skardal2020higher} which minimize the pairwise and triadic phase differences between the swarmalators, and the associated distance-dependent cosine terms amplify the coupling intensity among the interacting swarmalators. On the other hand, Eq. \eqref{sw_position_eq} serves as a counterpart of Eq. \eqref{sw_phase_eq} and captures phase-dependent swarming behavior. In this case, the sine terms minimize the distances between the swarmalators, leading them to aggregate or swarm together, while the cosine terms, similar to Eq. \eqref{sw_phase_eq}, bolster the coupling between the swarmalators, but this time based on their phase similarity. One can also interpret both Eqs. \eqref{sw_position_eq} and \eqref{sw_phase_eq} as models that represent synchronization on the unit torus with both pairwise and three-body interactions. Hence, the model provides a more general framework for swarmalators by considering beyond pairwise interactions.      

\par  In order to simplify the model, we convert the trigonometric function to complex exponentials and introduce new variables $\xi_{i}=x_{i}+\theta_{i}$ and $\eta_{i}=x_{i}-\theta_{i}$, which eventually provide,
\begin{subequations}  \label{xi_eta_eq}
 \begin{multline} \label{xi_eq}
    \dot{\xi}_{i} = v_{i}+ \omega_{i} + \dfrac{1}{2\mathrm{i}} [H_{1}^{+} e^{-\mathrm{i}\xi_{i}}- (H_{1}^{+})^{*}e^{\mathrm{i}\xi_{i}}] 
    \\ + \dfrac{1}{2\mathrm{i}} [H_{2}^{-} e^{-\mathrm{i}\eta_{i}}- (H_{2}^{-})^{*}e^{\mathrm{i}\eta_{i}}],
\end{multline}
\begin{multline}  \label{eta_eq}
    \dot{\eta}_{i} = v_{i}-\omega_{i} + \dfrac{1}{2\mathrm{i}} [H_{1}^{-} e^{\mathrm{i}\xi_{i}}- (H_{1}^{-})^{*}e^{\mathrm{i}\xi_{i}}] 
    \\ + \dfrac{1}{2\mathrm{i}} [H_{2}^{+} e^{-\mathrm{i}\eta_{i}}- (H_{2}^{+})^{*}e^{\mathrm{i}\eta_{i}}], 
\end{multline}  
\end{subequations}
where $\mathrm{i}=\sqrt{-1}$ and 
\begin{equation} \label{H12_eq}
    \begin{array}{l}
      H_{1}^{\pm}=J_{1}^{\pm} Z_{1}^{+} +J_{2}^{\pm} Z_{2}^{+} (Z_{1}^{+})^{*}, \\\\
       H_{2}^{\pm}=J_{1}^{\pm} Z_{1}^{-} +J_{2}^{\pm} Z_{2}^{-} (Z_{1}^{-})^{*}, 
    \end{array}
\end{equation}
with $J_{m}^{\pm}=\dfrac{J_{m}\pm K_{m}}{2}$ $(m=1,2)$ and 
\begin{equation}\label{order_para}
    \begin{array}{l}
     Z_{m}^{\pm}=\sum\limits_{j=1}^{N} e^{m\mathrm{i}(x_{j}\pm \theta_{j})}=S_{m}^{\pm}e^{\mathrm{i}\psi_{m}^{\pm}}.
    \end{array}
\end{equation}
Here $Z_{1}^{\pm} (S_{1}^{\pm})$ indicates the order parameters associated with the conventional swarmalator model that quantifies the space-phase order of the system \cite{o2017oscillators,o2022collective,yoon2022sync}. When the correlation between phase and space is perfect (i.e., $x_{i}=\pm \theta_{i}+C_{0}$, for some constant $C_{0}$), the value of the order parameter $S_{1}^{\pm}$ is equal to $1$. While, when $\theta_{i}$ and $x_{i}$ are uncorrelated, the value of $S_{1}^{\pm}$ is $0$. Therefore the order parameters $S_{1}^{\pm}$ measure the degree of correlation between the space $(x_{i})$ and phase $(\theta_{i})$ variables, with $S_{1}^{\pm}$ ranging from $0$ (no correlation) to $1$ (perfect correlation). On the other hand, $Z_{2}^{\pm} (S_{2}^{\pm})$ can be interpreted as new order parameters that come up as a result of higher-order interactions, analogous to the higher-order Kuramoto phase models \cite{skardal2020higher,skardal2019abrupt}. In our present study, we focus only on the evolution of conventional order parameters $Z_{1}^{\pm} (S_{1}^{\pm})$. 
\section{Results} \label{results}
\begin{figure*}[htp]
    \centerline{
    \includegraphics[scale=0.25]{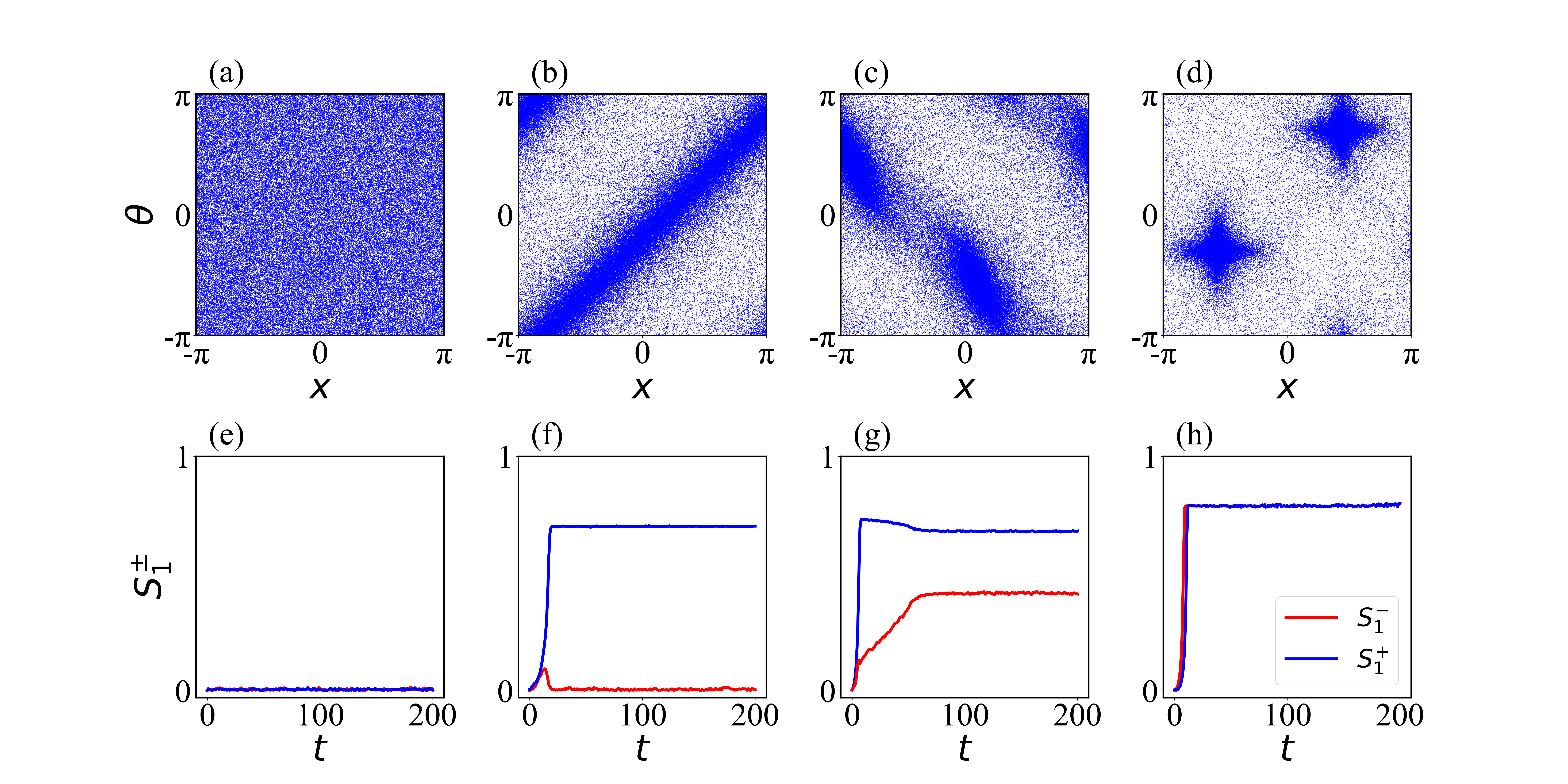}}
    \caption{{\bf Distinct collective states.} (a)-(d) Scatter plots of all the four states in $(x,\theta)$ plane. (e)-(f) Time evolution of order parameters $S_{1}^{+}(S_{1}^{-})$, depicted in blue (red). (a) and (e) represent the async state for a choice of coupling strengths $(J_{1}, K_{1}, J_{2}, K_{2})=(7,-1,8,9)$, (b) and (f) correspond to the phase wave state with $(J_{1}, K_{1}, J_{2}, K_{2})=(1,6.5,5,9)$, (c) and (g) are associated with the mixed state for $(J_{1}, K_{1}, J_{2}, K_{2})=(9,1.8,6.5,5.5)$, and the sync state is delineated in (d), (h) for $(J_{1}, K_{1}, J_{2}, K_{2})=(9,5,6.5,5.5)$. All the results are generated by integrating Eqs. \eqref{sw_position_eq} and \eqref{sw_phase_eq} using Julia Tsit5 adaptive differential equation solver \cite{rackauckas2017differentialequations} with $N=10^{5}$ swarmalators whose intrinsic velocity and frequency have been drawn from the Lorentzian distribution with zero mean and half-width $\Delta_{\omega}=\Delta_{v}=1$.}
    \label{all_states}
\end{figure*}
\par Now, the coupling dependency of order parameters $Z_{1}^{\pm} (S_{1}^{\pm})$ in Eqs. \eqref{xi_eta_eq} and \eqref{order_para} indicate that depending on the values of coupling strengths, several combinations for order parameters $(S_{1}^{+}, S_{1}^{-})$ can be achieved, which eventually leads to the emergence of different collective states. We, therefore, start by integrating the Eqs. \eqref{sw_position_eq} and \eqref{sw_phase_eq} for the calculation of the order parameters $S_{1}^{+}, S_{1}^{-}$ with $N=10^{5}$ swarmalators and the half-widths of Lorentzian distribution $\Delta_{v}=\Delta_{\omega}=1$. The results show that depending on the values of coupling strengths, four distinct stable states emerge in the system, which can be classified by the dyad $(S_{1}^{+}, S_{1}^{-})$ as follows:
\begin{enumerate}[(i)]
    \item The first state is referred to as the \qq{Async} state, denoted by $(S_{1}^{+},S_{1}^{-})=(0,0)$ state. In this state, the swarmalators are uniformly distributed in phase and space, as demonstrated in Fig.~\ref{all_states}(a). There is no space-phase order among the swarmalators and so $(S_{1}^{+}, S_{1}^{-}) \approx (0,0)$ [see Fig.~\ref{all_states}(e)].
    \item \qq{Phase waves} or $(S,0)$ or $(0, S)$ state [Figs.~\ref{all_states}(b) and \ref{all_states}(f)]: In this state the swarmalators develop a band or phase wave, where the spatial positions $x_{i}$ and phase angles $\theta_{i}$ are related as $x_{i} \approx \mp \theta_{i}$, depending on whether it is $(S,0)$ or $(0, S)$ state, respectively. In the $(\xi,\eta)$ coordinate system, the swarmalators are partially locked in either $\xi_{i}$ or $\eta_{i}$ and drift in the other variable.
    \item The third state is referred to as the \qq{Mixed state}, denoted by $(S_{1}^{+},S_{1}^{-})=(S', S'')$, where $S' \neq S'' \neq 0$ (Fig.~\ref{all_states}(g)). In this state, the swarmalators form a band where clusters of correlated swarmalators are observed to move together, as depicted in Fig.~\ref{all_states}(c).
    \item The fourth state is known as \qq{Sync} state, denoted by $(S,S)$, where $S \neq 0$ [Fig. \ref{all_states}(h)]. In this state, the swarmalators are partially locked in both $\xi_{i}$ and $\eta_{i}$. For most initial conditions, two clusters of locked swarmalators emerge spontaneously, as depicted in Fig.~\ref{all_states}(d). However, one can also observe a single cluster of locked swarmalators for some initial conditions.   
\end{enumerate}
Next, in order to analyze all these four states, we employ the Ott-Antonsen (OA) ansatz \cite{ott2008low} in the thermodynamic $(N \to \infty)$ limit and derive the expressions for order parameters in each state. In the $N \to \infty$ limit, the collective states of the swarmalators can be defined by a continuous function $\rho(v,\omega,\xi,\eta,t)$ as,
\begin{equation} \label{OA_density_func}
    \begin{array}{l}
        \rho \equiv \dfrac{1}{N} \sum\limits_{j=1}^{N} \delta(v-v_{j}) \delta(\omega-\omega_{j}) \delta(\xi-\xi_{j}) \delta(\eta-\eta_{j}),
    \end{array}
\end{equation}
where $\rho(v,\omega,\xi,\eta,t)$ is the probability to have a swarmalator at time $t$ with intrinsic frequency $\omega$, intrinsic velocity $v$, and coordinates $\eta$ and $\xi$.  
Differentiating \eqref{OA_density_func} with respect to $t$, one can obtain the continuity equation 
\begin{equation} \label{continuity_eq}
    \begin{array}{l}
          \dfrac{\partial \rho}{\partial t} + \dfrac{\partial}{\partial \xi} (\dot{\xi}\rho) + \dfrac{\partial}{\partial \eta} (\dot{\eta}\rho) = 0.
    \end{array}
\end{equation}
Given that our model is a higher-order Kuramoto model occurring on a torus, we are in pursuit of a \qq{torodoidal} OA ansatz \cite{ott2009long,ott2008low,yoon2022sync}, which can be described as a multiplication of Poisson kernels,
\begin{multline} \label{density_func_poison_kernel}
          \rho(v,\omega,\xi,\eta,t)=\dfrac{1}{4\pi^{2}}g_{v}(v)g_{\omega}(\omega)\biggl[ 1+ \sum\limits_{p=1}^{\infty} \alpha^{p} e^{\mathrm{i}p\xi}+ \mbox{c.c} \biggr]\\ 
        \times  \biggl[ 1+ \sum\limits_{q=1}^{\infty} \beta^{q} e^{\mathrm{i}q\eta}+ \mbox{c.c} \biggr],
\end{multline}
where $\alpha(v,\omega,t)$ and $\beta(v,\omega,t)$ are undetermined and need to be solved in a self-consistent manner, and \qq{c.c} refers to the complex conjugate of its preceding terms. Now plugging the expression for $\rho$ [given by Eq.~\eqref{density_func_poison_kernel}] into the continuity Eq.~\eqref{continuity_eq}, we obtain that the Fourier modes $\alpha(v,\omega,t)$ and $\beta(v,\omega,t)$ are subsequently constrained to adhere to the identical conditions for all harmonics $p$ and $q$, which leads to the fulfillment of a coupled complex-valued differential equation as follows, 
\begin{subequations}\label{eq_alpha_beta}
\begin{multline} \label{alpha_eq}
        \dot{\alpha} = - \mathrm{i} (v+\omega) \alpha + \dfrac{1}{2} [(H_{1}^{+})^{*}-H_{1}^{+}\alpha^{2}] \\
      + \dfrac{\alpha}{2} [(H_{2}^{-})^{*}\beta^{*}-H_{2}^{-}\beta],
\end{multline}
\begin{multline} \label{beta_eq}
        \dot{\beta} = - \mathrm{i} (v-\omega) \beta + \dfrac{1}{2} [(H_{2}^{+})^{*}-H_{2}^{+}\beta^{2}] \\
      + \dfrac{\beta}{2} [(H_{1}^{-})^{*}\alpha^{*}-H_{1}^{-}\alpha],
\end{multline}
\end{subequations}
in the submanifold $\norm{\alpha}=1=\norm{\beta}$. Subsequently, the order parameters $Z_{m}^{\pm}$ $(m=1,2)$ become
\begin{subequations}
    \begin{equation} \label{alpha_order_para} 
    \begin{array}{l}
         Z_{m}^{+} = \int_{-\infty}^{\infty} dv  \int_{-\infty}^{\infty} d\omega g_{v}(v) g_{\omega}(\omega) \alpha^{*^{m}}(v,\omega,t),
    \end{array}
    \end{equation}
      \begin{equation}\label{beta_order_para}
    \begin{array}{l}
         Z_{m}^{-} = \int_{-\infty}^{\infty} dv  \int_{-\infty}^{\infty} d\omega g_{v}(v) g_{\omega}(\omega) \beta^{*^{m}}(v,\omega,t).
    \end{array}
    \end{equation}   
\end{subequations}
Equations \eqref{alpha_eq}-\eqref{beta_order_para} contain a set of self-consisting equations for the order parameters $Z_{m}^{\pm}$ in the $N\to \infty$ limit.

\subsection{Stability of the async state} \label{async}
In the async state, the order parameters $Z_{m}^{\pm}$ are zero. When $Z_{m}^{\pm}=0$, Eqs. \eqref{alpha_eq} and \eqref{beta_eq} give the solutions $\alpha_{0}(v,\omega,t)=\exp{[-\mathrm{i}(v+\omega)t]}$ and  $\beta_{0}(v,\omega,t)=\exp{[-\mathrm{i}(v-\omega)t]}$. Clearly, these solutions are self-consistent, as substituting them into the Eqs. \eqref{alpha_order_para} and \eqref{beta_order_para} one can obtain $Z_{m}^{\pm}=\exp{[-\mathrm{i}(\Delta_{v}+\Delta_{\omega})t]}$, which converges to zero in the $t \to \infty$ limit.
\par Now, to investigate the stability of the async state, we introduce a small perturbation around the solutions $\alpha_{0}$ and $\beta_{0}$ given by
\begin{equation}\label{async_perturbation}
    \begin{array}{l}
         \alpha_{1}(v,\omega,t)=\alpha(v,\omega,t)-\alpha_{0}(v,\omega,t), \\\\
         \beta_{1}(v,\omega,t)=\beta(v,\omega,t)-\beta_{0}(v,\omega,t).
    \end{array}
\end{equation}
This eventually gives the perturbed order parameters as
\begin{equation} \label{perturbed_order_para}
    \begin{array}{l}
        Z_{11}^{+} = \int_{-\infty}^{\infty} dv  \int_{-\infty}^{\infty} d\omega g_{v}(v) g_{\omega}(\omega) \alpha_{1}^{*}(v,\omega,t), \\\\
        Z_{11}^{-} = \int_{-\infty}^{\infty} dv  \int_{-\infty}^{\infty} d\omega g_{v}(v) g_{\omega}(\omega) \beta_{1}^{*}(v,\omega,t),
    \end{array}
\end{equation}
where we assume that $1 \gg \norm{\alpha_{1}(v,\omega,t)}$, $\norm{\beta_{1}(v,\omega,t)}$ and $\norm{Z_{11}^{\pm}}$. Substituting the expressions for $\alpha_{1}$, $\beta_{1}$ and $Z_{11}^{\pm}$ into the Eqs. \eqref{alpha_eq} and \eqref{beta_eq}, and considering the terms up to first order, we obtain the following set of evolution equations 
\begin{subequations} 
\begin{multline} \label{alpha1_eq}
        \dot{\alpha}_{1} = - \mathrm{i} (v+\omega) \alpha_{1} + \dfrac{1}{2} [J_{1}^{+}(Z_{11}^{+})^{*}-J_{1}^{+}Z_{11}^{+}\alpha_{0}^{2}] 
        \\ + \dfrac{\alpha_{0}}{2\beta_{0}} [J_{1}^{-}(Z_{11}^{-})^{*}-J_{1}^{-}Z_{11}^{-}\beta_{0}^{2}],
\end{multline}
\begin{multline}\label{beta1_eq}
        \dot{\beta}_{1} = - \mathrm{i} (v-\omega) \beta_{1} + \dfrac{1}{2} [J_{1}^{+}(Z_{11}^{-})^{*}-J_{1}^{+}Z_{11}^{-}\beta_{0}^{2}] 
        \\ + \dfrac{\beta_{0}}{2\alpha_{0}} [J_{1}^{-}(Z_{11}^{+})^{*}-J_{1}^{-}Z_{11}^{+}\alpha_{0}^{2}].
\end{multline}
\end{subequations}
One can notice that the terms with $\alpha_{0}$ and $\beta_{0}$ as a function of $v$ and $\omega$ oscillate rapidly for $t \gg 1$. These rapidly oscillating terms barely contribute when integrating over $v$ and $\omega$, and thus to approximate the integrals, we can neglect these small terms. Now, introducing a new variable $\tau=v+\omega$, $\alpha$ and $\beta$ can be expressed as $\alpha(v,\omega,t)=\alpha(y,t)$, and $\beta(v,\omega,t)=\beta(y,t)$, respectively. Then, integrating the Eqs. \eqref{alpha1_eq} and \eqref{beta1_eq} over $v$ and $\omega$, we eventually obtain  
\begin{widetext}
\begin{subequations}
    \begin{equation}\label{perturbed_order_para2}
    \begin{array}{l}
         \dfrac{dZ_{11}^{+}}{dt} = \int_{-\infty}^{\infty} dv  \int_{-\infty}^{\infty} d\omega g_{v}(v) g_{\omega}(\omega) \dfrac{d\alpha_{1}^{*}}{dt} = \int_{-\infty}^{\infty} d\tau \mathcal{G}(\tau) \dfrac{d\alpha_{1}^{*}(\tau)}{dt} = \dfrac{J_{1}^{+}-2(\Delta_{v}+\Delta_{\omega})}{2}Z_{11}^{+},
    \end{array}
    \end{equation}
      \begin{equation} \label{perturbed_order_para3}
    \begin{array}{l}
        \dfrac{dZ_{11}^{-}}{dt} = \int_{-\infty}^{\infty} dv  \int_{-\infty}^{\infty} d\omega g_{v}(v) g_{\omega}(\omega) \dfrac{d\beta_{1}^{*}}{dt} = \int_{-\infty}^{\infty} d\tau \mathcal{G}(\tau) \dfrac{d\beta_{1}^{*}(\tau)}{dt} = \dfrac{J_{1}^{+}-2(\Delta_{v}+\Delta_{\omega})}{2}Z_{11}^{-},
    \end{array}
    \end{equation} 
\end{subequations}
where 
\begin{equation}
    \begin{array}{l}
       \mathcal{G}(\tau)= \int_{-\infty}^{\infty} dv  \int_{-\infty}^{\infty} d\omega g_{v}(v) g_{\omega}(\omega) \delta[\tau-(v\pm \omega)]= \dfrac{\Delta_{v}+\Delta_{\omega}}{\pi[\tau^{2}+(\Delta_{v}+\Delta_{\omega})^{2}]}.
    \end{array}
\end{equation}
\end{widetext}
To evaluate the integration, we use the fact that it has a residue $\tau=\mathrm{i}(\Delta_{v}+\Delta_{\omega})$ in the upper half plane where $\alpha_{0}^{*}(\tau)$ and $\beta_{0}^{*}(\tau)$ are analytic. Now, the async state becomes stable when the perturbed order parameters $Z_{11}^{\pm}$ die out in time. Hence, from Eqs. \eqref{perturbed_order_para2} and \eqref{perturbed_order_para3}, one can conclude that the async state becomes unstable when $J_{1}^{+}-2(\Delta_{v}+\Delta_{\omega}) >0$, or in other words the async state sustains its stability for $J_{1}^{+}-2(\Delta_{v}+\Delta_{\omega})<0$. Therefore, the curve satisfying 
\begin{equation}\label{async_critical}
\begin{array}{l}
    J_{1}^{+}=2(\Delta_{v}+\Delta_{\omega}),
    \end{array}
\end{equation}
signifies the critical curve above which the async state loses its stability, and the swarmalators form a phase wave state. Eq. \eqref{async_critical} reveals that the transition from async state to phase wave state depends solely on the pairwise interactions by means of the pairwise coupling strengths $(J_{1}, K_{1})$.

\subsection{Analysis of phase wave state} \label{phase_wave}
In the phase wave state, swarmalators develop a phase wave or band with $x_{i}=\mp \theta_{i}$ for $(S,0)$ and $(0, S)$ states, respectively. Therefore, we seek a solution to Eqs. \eqref{alpha_eq}-\eqref{beta_order_para} that satisfies $\dot{\alpha}=0$, $\dot{\beta}\neq 0$, $Z_{m}^{+} \neq 0$, and $Z_{m}^{-} = 0$, $(m=1,2)$. Substituting these relations into the Eqs. \eqref{alpha_eq} and \eqref{beta_eq}, we have 
\begin{subequations}
\begin{equation}\label{phase_alpha_dot_eq}
    \begin{array}{l}
        0 = - \mathrm{i} (v+\omega) \alpha + \dfrac{1}{2} [(H_{1}^{+})^{*}-H_{1}^{+}\alpha^{2}],
    \end{array}
\end{equation}
\begin{equation}\label{phase_beta_dot_eq}
    \begin{array}{l}
        \dot{\beta} = - \mathrm{i} (v-\omega) \beta + \dfrac{\beta}{2\alpha} [(H_{1}^{-})^{*}-H_{1}^{-}\alpha^{2}].
    \end{array}
\end{equation}
\end{subequations}
Notice that $\alpha(v,\omega,t)$ depends on $v+\omega$, which is distributed according to a Lorentzian distribution with spread $\Delta_{v}+\Delta_{\omega}$. This allows us to evaluate the integral for the order parameter $Z_{m}^{+}$ explicitly as $Z_{1}^{+}=\alpha^{*}(i\Delta_{v}+i\Delta_{\omega},t)$ using the Cauchy's residue theorem by closing the contour to an infinite-radius semicircle in the upper half-plane. Similarly, we can obtain $ Z_{2}^{+}=\alpha^{{*}^{2}}(i\Delta_{v}+i\Delta_{\omega},t)=(Z_{1}^{+})^{2} $. Substituting the relation between $Z_{1}^{+}$ and $Z_{2}^{+}$ into the Eqs. \eqref{phase_alpha_dot_eq} and \eqref{phase_beta_dot_eq}, and assuming $\psi_{1}^{+}=0$, we obtain the expressions for $\alpha$ and $\beta$ as follows,
\begin{equation}\label{phase_alpha_eq}
    \begin{array}{l}
         \alpha(v,\omega)=\mathcal{F}\biggl(\dfrac{v+\omega}{S_{1}^{+}(J_{1}^{+}+(S_{1}^{+})^{2}J_{2}^{+})} \biggr),
    \end{array}
\end{equation}
and
\begin{equation}\label{phase_beta_eq}
    \begin{array}{l}
         \beta(v,\omega,t)= e^{(-rt)},
    \end{array}
\end{equation}
where we introduce a function $\mathcal{F}$ as 
\begin{equation} \label{F_function}
    \begin{array}{l}
         \mathcal{F}(y)=-\mathrm{i}y + \sqrt{1-y^{2}},
    \end{array}
\end{equation}
and the term $r$ is given by,
\begin{widetext}
\begin{equation}\label{r_eq}
    \begin{array}{l}
         r= \mathrm{i}(v-\omega) -\dfrac{1}{2\alpha}[(H_{1}^{-})^{*}-H_{1}^{-}\alpha^{2}] = \mathrm{i}\biggl[ \dfrac{J_{1}K_{1}}{J_{1}^{+}+(S_{1}^{+})^{2}J_{2}^{+}} \biggl(\dfrac{v}{J_{1}}-\dfrac{\omega}{K_{1}}\biggr) + \dfrac{(S_{1}^{+})^{2}J_{2}K_{2}}{J_{1}^{+}+(S_{1}^{+})^{2}J_{2}^{+}} \biggl(\dfrac{v}{J_{2}}-\dfrac{\omega}{K_{2}}\biggr) \biggr].
    \end{array}
\end{equation}
\end{widetext}
Equation \eqref{phase_beta_eq} results in $Z_{m}^{-}=0$, as anticipated. On the other hand, Eq. \eqref{phase_alpha_eq} implies that
\begin{equation}\label{phase_s_plus}
    \begin{array}{l}
         S_{1}^{+}=\mathcal{F}^{*}\biggl(\dfrac{\mathrm{i}\Delta_{v}+\mathrm{i}\Delta_{\omega}}{S_{1}^{+}(J_{1}^{+}+(S_{1}^{+})^{2}J_{2}^{+})} \biggr).
    \end{array}
\end{equation}
Solving Eq. \eqref{phase_s_plus} for $S_{1}^{+}$ gives the expression of $S_{1}^{+}$ as
\begin{equation} \label{phase_wave_S+}
    \begin{array}{l}
      S_{1}^{+}= {\sqrt{\frac{J_{2}^{+}-J_{1}^{+}\pm \sqrt{\left(J_{1}^{+}+J_{2}^{+}\right)^{2}-8 J_{2}^{+} \tilde{\Delta }}}{2 J_{2}^{+}}}},
    \end{array}
\end{equation}
where $\tilde{\Delta}=\Delta_{v}+\Delta_{\omega}$. The plus and minus sign inside the square root corresponds to the stable and unstable solutions if they exist. Equation~\eqref{phase_wave_S+} suggests the presence of a bistable region upon the selection of coupling strengths. In this scenario, a particular combination of coupling values leads to the coexistence of asynchronous and phase wave states depending on the initial conditions chosen. Saying differently, with varying coupling strengths, there exist two different transition scenarios: one corresponds to the transition from the async state to the phase wave state, referred to as forward transition, and the other is backward transition, which is associated with the transition from phase wave state to async state. For the forward transition case, $S_{1}^{+}$ bifurcates from $0$ (i.e., async state) at
\begin{equation}\label{phase_fwd_critical}
    \begin{array}{l}
         J_{1,f}^{+}=2\tilde{\Delta}=2(\Delta_{v}+\Delta_{\omega}), 
    \end{array}
\end{equation}
which is consistent with Eq. \eqref{async_critical}, where forward transition from async state to phase wave state emerges and only the stable branch of $S_{1}^{+}$ exists. This once again guarantees that the transition from async state to phase wave state is solely dependent on the pairwise coupling strengths. On the other hand, in the case of backward transition, both the stable and unstable branches coexist for a region of parameter values called the hysteresis region. But as soon as the backward critical coupling condition $J_{1}^{+}=J_{1,b}^{+}$ is reached, both the stable and unstable branches clash and destroy each other. As a result, the stability of the phase wave is completely shattered, and $S_{1}^{+}=0$ is the sole viable solution. From Eq. \eqref{phase_wave_S+}, we obtain that both the stable and unstable branches of $S_{1}^{+}$ exist if the following coupling condition satisfies, 
\begin{equation}\label{phase_bwd_critical}
    \begin{array}{l}
         J_{1,b}^{+}=2\sqrt{2J_{2}^{+}\tilde{\Delta}}-J_{2}^{+}. 
    \end{array}
\end{equation}
Eq. \eqref{phase_bwd_critical} thus corresponds to the critical curve for the backward transition. 
\par
\begin{figure}[htp]
    \includegraphics[width=\columnwidth]{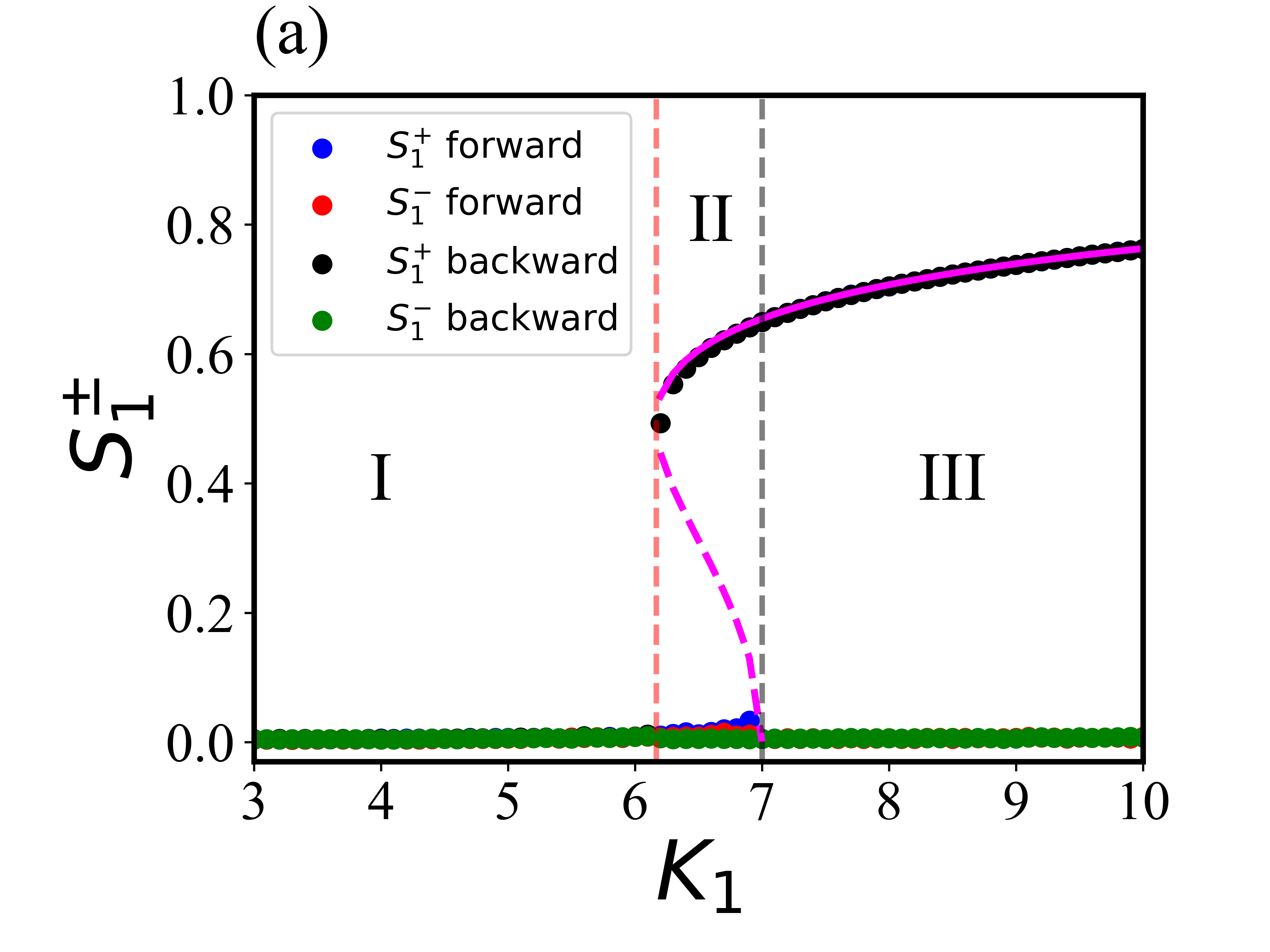}
    \includegraphics[width=\columnwidth]{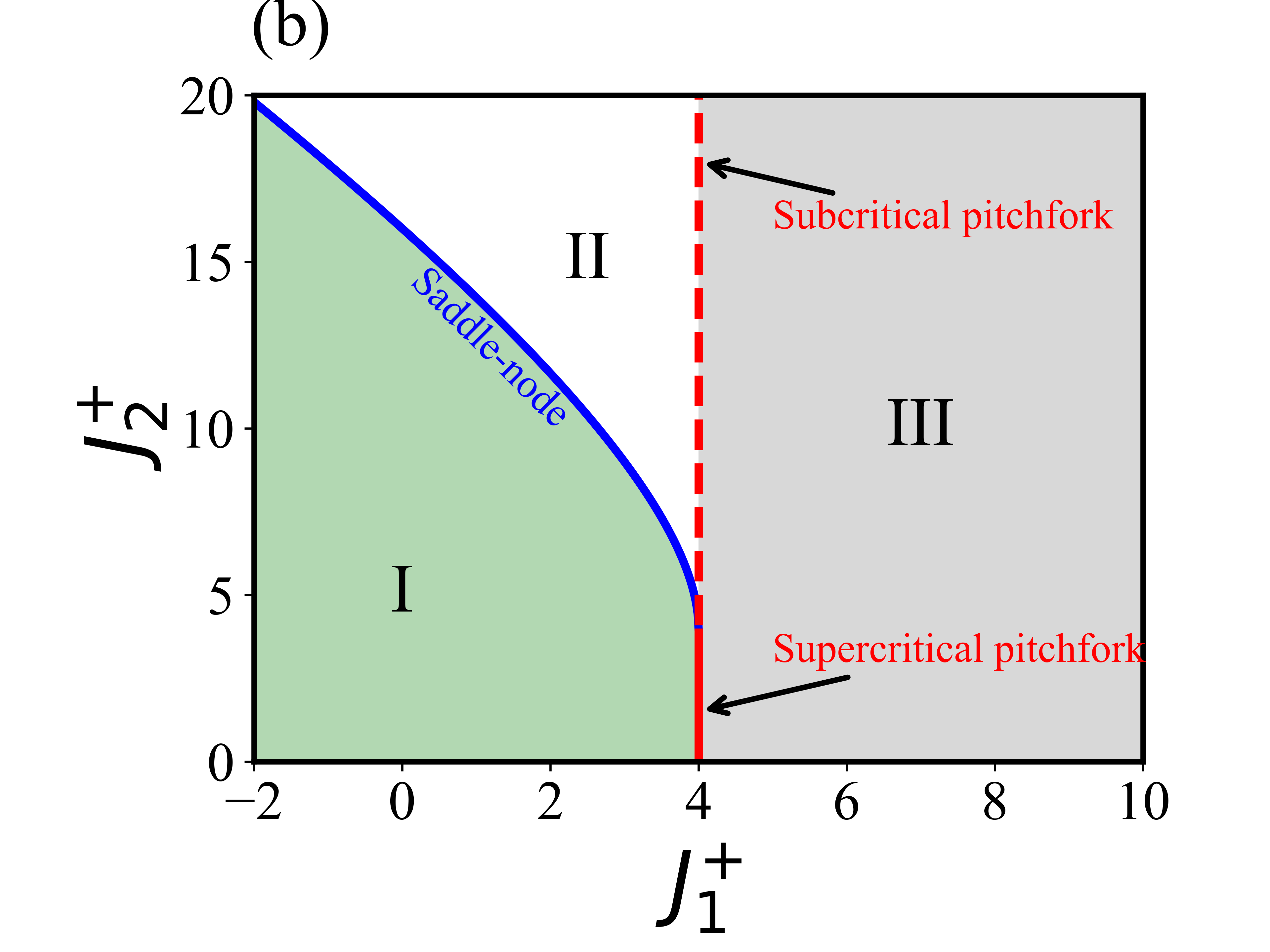}
    \caption{{\bf Abrupt transition from the async state to the phase wave state.} (a) Order parameter $S_{1}^{\pm}$ as a function of pairwise coupling strength $K_{1}$ for $J_{1}=1$ and fixed three-body coupling strengths $J_{2}=5$, $K_{2}=9$. Solid and dashed magenta curves represent the stable and unstable solutions given by Eq.\eqref{phase_wave_S+}, respectively. The dashed vertical lines correspond to the critical couplings for forward (in red) and backward (in black) transitions obtained from Eqs. \eqref{phase_fwd_critical} and \eqref{phase_bwd_critical}, respectively. Solid circles depict the result obtained from the direct simulation of Eqs. \eqref{sw_position_eq} and \eqref{sw_phase_eq} for $N=10^{5}$ oscillators with half widths of the Lorentzian distribution $\Delta_{v}=\Delta_{\omega}=1$. The results reveal an abrupt transition from async $(I)$ state to phase wave $(III)$ state, which results in a bistable domain $(II)$ where both the referred states are stable. (b) The comprehensive stability diagram illustrating the states of async $(I)$, phase wave $(III)$ and bistability $(II)$ as a function of pairwise coupling $J_{1}^{+}=\frac{J_{1}+K_{1}}{2}$ and higher-order coupling $J_{2}^{+}=\frac{J_{2}+K_{2}}{2}$. Two distinct types of bifurcations, saddle-node and pitchfork, are represented by blue and red curves, respectively. These two curves intersect each other at $(J_{1}^{+},J_{2}^{+})=(4,4)$. When $J_{2}^{+}<4$, the pitchfork bifurcation is deemed supercritical, whereas when $J_{2}^{+}>4$, the pitchfork bifurcation becomes subcritical.}
    \label{phase_wave_fwd_bwd}
\end{figure}

\begin{figure}
    \centerline{
    \includegraphics[scale=0.26]{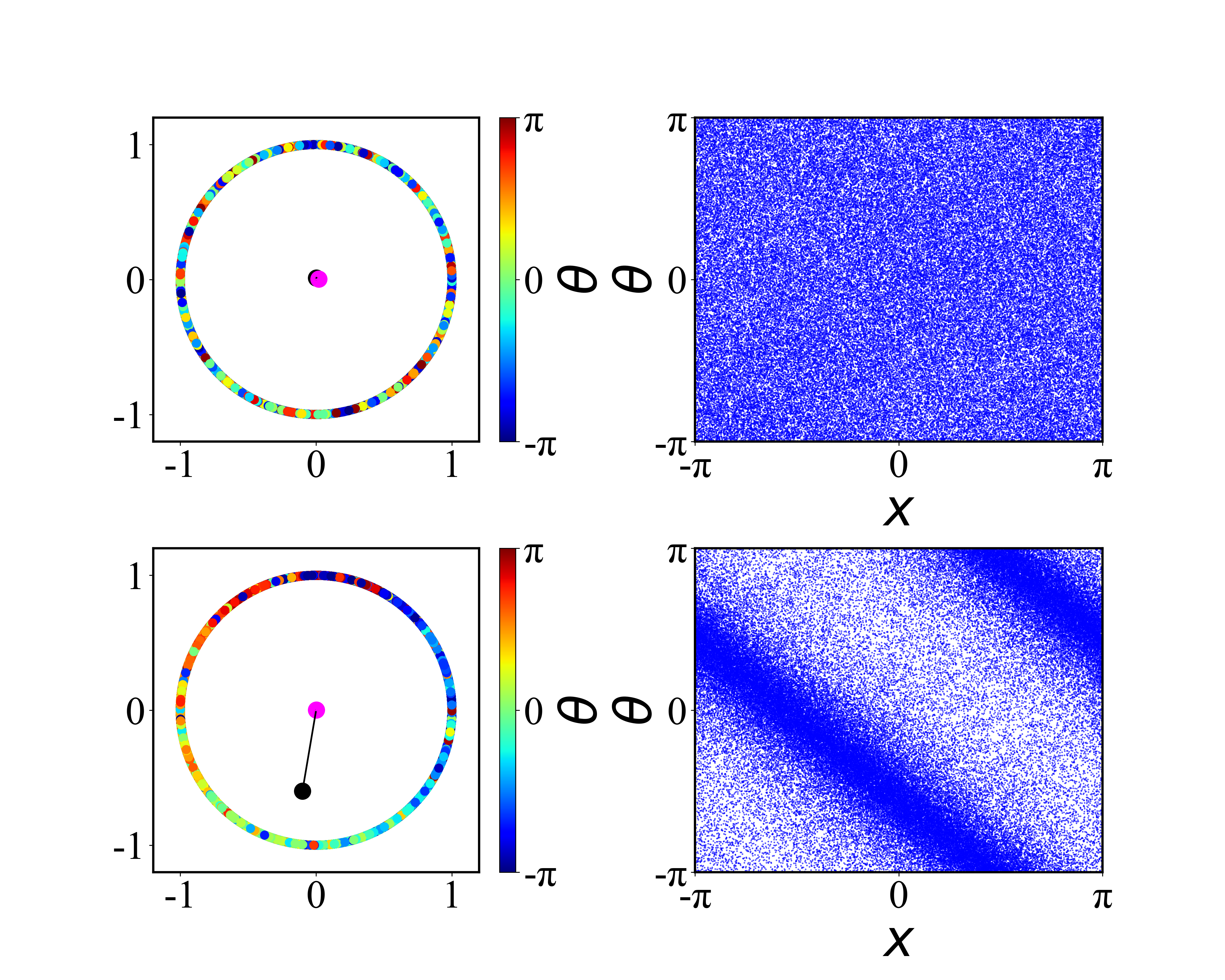}}
    \caption{{\bf Bistablilty between async and phase wave state}. Scatter plot for async $(0,0)$ and phase wave $(S,0)$ states at $K_{1}=6.5$, $J_{1}=1$, $J_{2}=5$, and $K_{2}=9$ [drawn from the region $(II)$ in Fig. \ref{phase_wave_fwd_bwd}(a)] is depicted in the upper and lower row, respectively. The order parameter $S_{1}^{+}(S_{1}^{-})$ is indicated by the black (magenta) circle in the left column, where the values of $S_{1}^{+}$ and $S_{1}^{-}$ are represented by the length of the line joining the center with the respective circles. Clearly, in the upper row, both the values of order parameters are almost zero, indicating the async $(0,0)$ state, while in the lower row, the value of $S_{1}^{+}$ is non-zero and $S_{1}^{-}$ is zero, characterizing the phase wave $(S,0)$ state. In the right panel, the corresponding scatter plots in the $(x,\theta)$ plane are displayed. Swarmalators are uniformly distributed in both phase and space, characterizing the async state (in the upper row), whereas in the lower row, swarmalators display a correlation between phase and space and thus correspond to the phase wave state.}
    \label{async_phase_snapshot}
\end{figure}

 To illustrate this, we numerically integrate the Eqs. \eqref{sw_position_eq} and \eqref{sw_phase_eq} with $N=10^{5}$ swarmalators for the calculation of $S_{1}^{\pm}$. In Fig. \ref{phase_wave_fwd_bwd}(a) we plot the variation of order parameter $S_{1}^{\pm}$ as a function of pairwise coupling strength $K_{1}$. $K_{1}$ is first increased adiabatically from $K_{1}=0$ to an adequately large value and then decreased back with the other couplings fixed at $J_{1}=1$, $J_{2}=5$, and $K_{2}=9$. The results reveal that $S_{1}^{-}$ remains zero all the time, while an abrupt transition from $S_{1}^{+} \approx 0$ to $S_{1}^{+} \approx 0.7$ occurs at $K_{1}=7$ [obtained from Eq. \eqref{phase_fwd_critical}] as the coupling strength $K_{1}$ is increased. Another abrupt transition from $S_{1}^{+} \approx 1$ to $S_{1}^{+} \approx 0$  occurs at $K_{1}=6$ [obtained from Eq. \eqref{phase_bwd_critical}] as $K_{1}$ is decreased starting from the phase wave state. Therefore, the system supports bistability behavior (where both phase wave and async states are stable) for $K_{1} \in [6,7]$. This is further substantiated by our theoretical predictions regarding the order parameters (illustrated using continuous and dashed magenta lines for the stable and unstable branches, respectively), demonstrating a commendable concurrence with the numerical findings (represented by solid circles). To emphasize the bistability nature, in Fig. \ref{async_phase_snapshot} we demonstrate the async (upper row) and phase wave (lower row) states, respectively, for $K_{1}=6.5$ while keeping the other couplings fixed at specific values as earlier. 
\par The outcomes given above highlight a significant finding that the introduction of higher-order interactions can cause a sudden transition from the async state to phase wave state and vice-versa, which was not the case with only pairwise interactions among the swarmalators \cite{yoon2022sync}. Thereafter, to better understand the effect of higher-order interactions in promoting bistability behavior, we plot the complete stability profile for the system in Fig.~\ref{phase_wave_fwd_bwd}(b). It shows that for adequately small values of higher-order coupling $(J_{2}^{+}<4)$, the transition from async state $(I)$ to phase wave state $(III)$ is continuous and takes place through a supercritical pitchfork bifurcation at $J_{1}^{+}=4$. However, for larger values of higher-order coupling $(J_{2}^{+}>4)$, the pitchfork bifurcation at $J_{1}^{+}=4$ becomes subcritical and a saddle-node bifurcation emerges at a lower value of $J_{1}^{+}$, given by Eq. \eqref{phase_bwd_critical} (depicted in blue). These two bifurcations correspond to the sudden transition observed in Fig. \ref{phase_wave_fwd_bwd}(a), and the region bounded by them refers to the region of bistability $(II)$ between async and phase wave state. Figure \ref{phase_wave_fwd_bwd}(b) also reveals an interesting observation that for relatively larger values of higher-order coupling $(J_{2}^{+}\ge 16)$, the region of bistability stretches into the negative region $J_{1}^{+}<0$, demonstrating the fact that higher-order interactions can stabilize the phase wave state even when the pairwise interactions are repulsive.     

\subsection{Analysis of sync state} \label{sync}
In the sync state we have $(S_{1}^{+},S_{1}^{-})=(S,S)$, with $S \neq 0$. Therefore, we seek solutions to the Eqs. \eqref{alpha_eq}-\eqref{beta_order_para} such that $Z_{m}^{\pm} \neq 0$ $(m=1,2)$. We will here analyze the sync state in two cases: one when $J_{1}=K_{1}$, $J_{2}=K_{2}$ and the other for an arbitrary combination of pairwise and higher-order couplings, i.e., for a generic case.  
\subsubsection{Specific case: $J_{1}=K_{1}$ and $J_{2}=K_{2}$}
\begin{figure}
    \centering
    \includegraphics[scale=0.5]{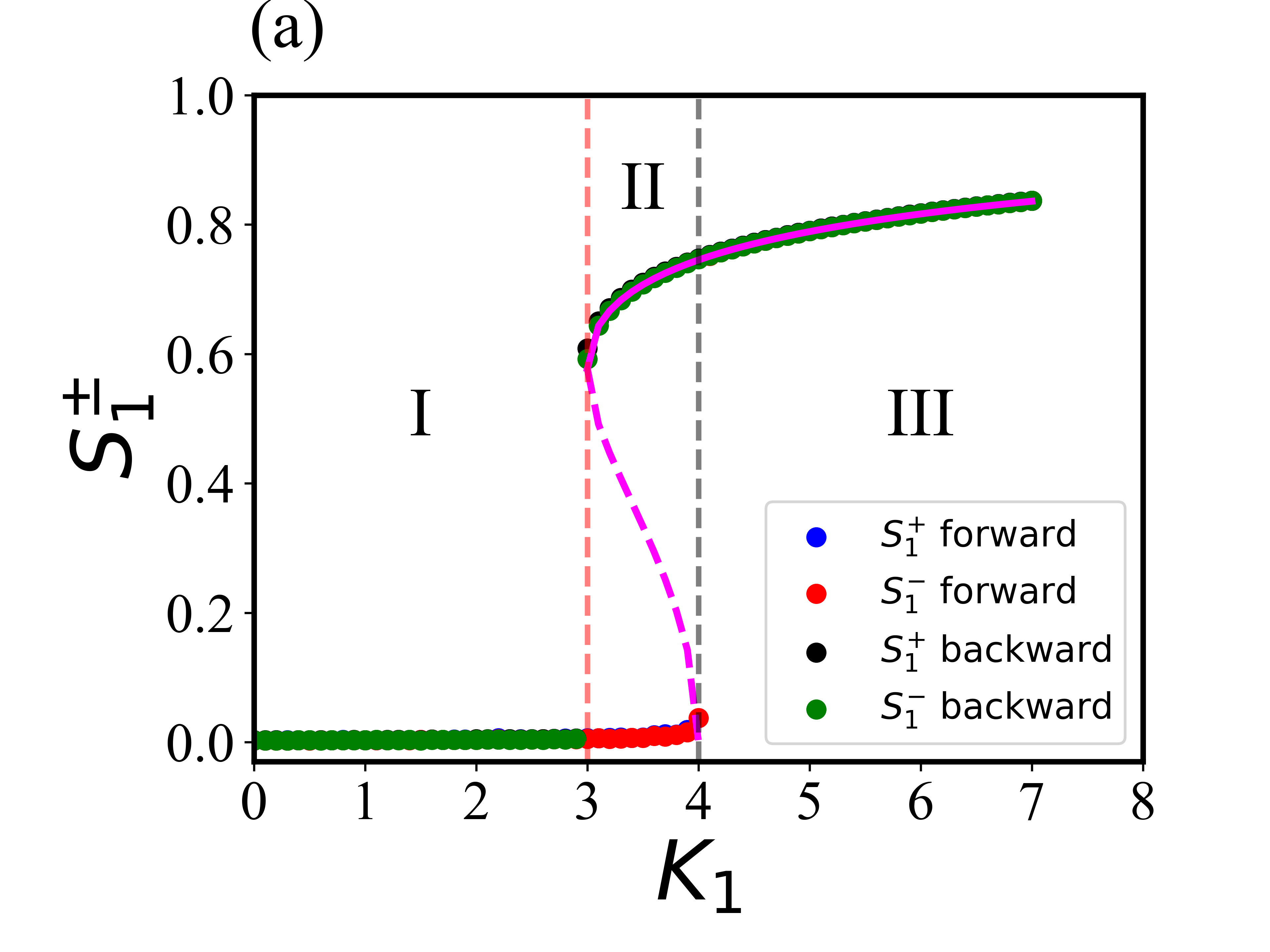}
    \includegraphics[scale=0.5]{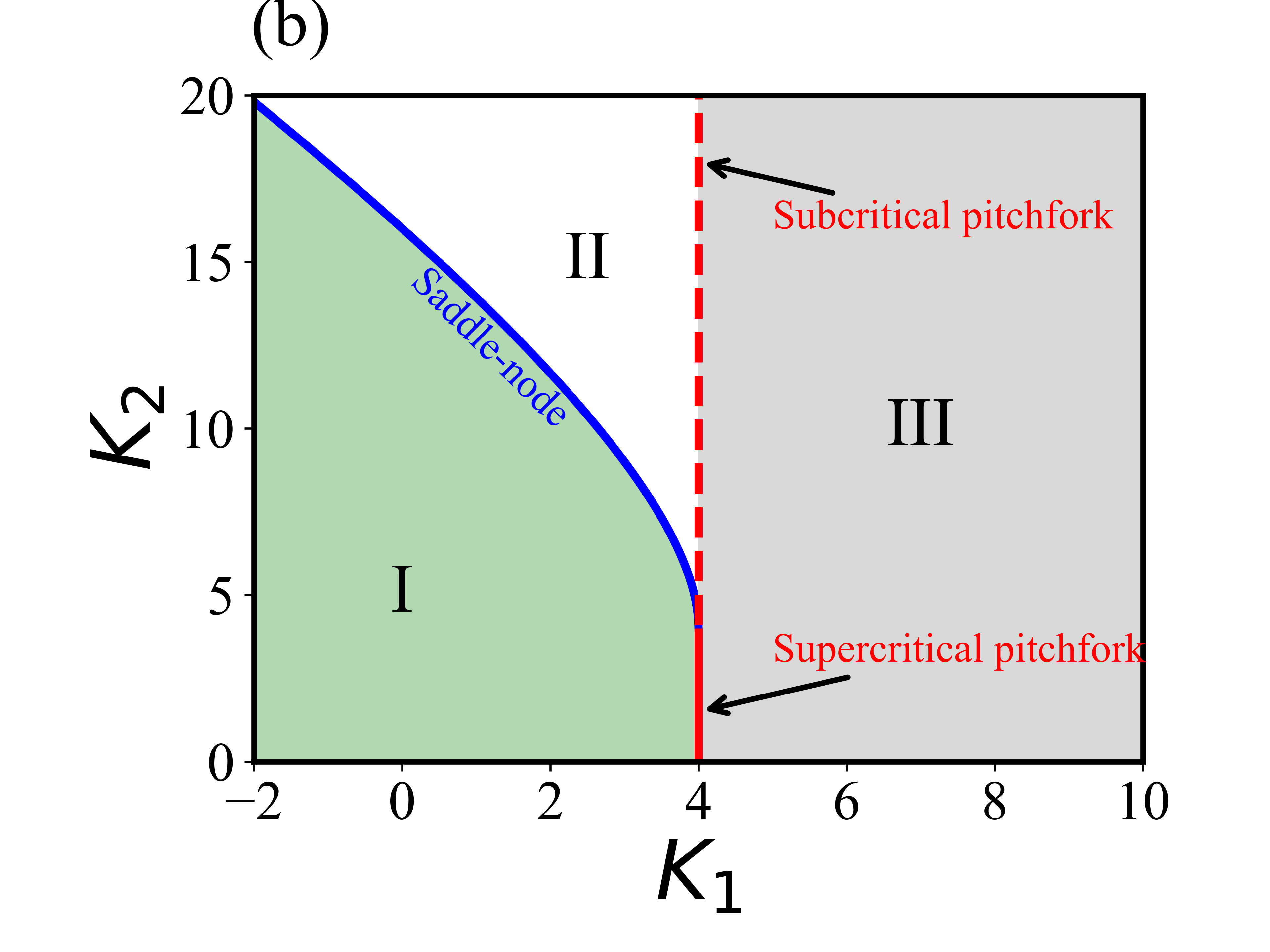}
    \caption{ {\bf Abrupt transition from async to sync state}. In (a), the behavior of the order parameter $S_{1}^{\pm}$ is presented in relation to the pairwise coupling strength $K_1$. With a fixed setting of $J_{1}=K_{1}$ and constant three-body coupling strengths $J_{2} = K_{2}=9$, both stable and unstable solutions (obtained from Eq. \eqref{J=K_sync_solution}) are showcased through solid and dashed magenta curves, respectively. The dashed vertical lines correspond to critical coupling values for forward (red) and backward (black) transitions. These values are derived from Eqs. \eqref{j=k_fwd_critical} and \eqref{phase_bwd_critical}, with $J_{1}=K_{1},J_{2}=K_{2}$. The outcome obtained from direct simulations of Eqs.\eqref{sw_position_eq} and \eqref{sw_phase_eq} is depicted in solid circles for $N =10^{5}$ swarmalators with Lorentzian distribution half-widths $\Delta_{\omega}=\Delta_{v}=1$. The findings distinctly reveal an abrupt transition from the async $(I)$ state to the sync $(III)$ state, establishing a bistable domain $(II)$ where both states are stable. Moving to (b), a comprehensive stability diagram is presented, illustrating the states of async $(I)$, phase wave $(III)$, and bistability $(II)$ as functions of pairwise coupling $K_{1}$ and higher-order coupling $K_{2}$. The diagram encompasses two distinct types of bifurcations: the blue curve representing a saddle-node bifurcation and the red curve depicting a pitchfork bifurcation. Notably, these curves intersect at $(K_{1},K_{2}) = (4, 4)$. When $K_{2}<4$, the pitchfork bifurcation takes a supercritical nature, whereas, for $K_{2}>4$, the pitchfork bifurcation becomes subcritical.}
    \label{j=k_fwd_bwd}
\end{figure}

\begin{figure}
    \centerline{
    \includegraphics[scale=0.26]{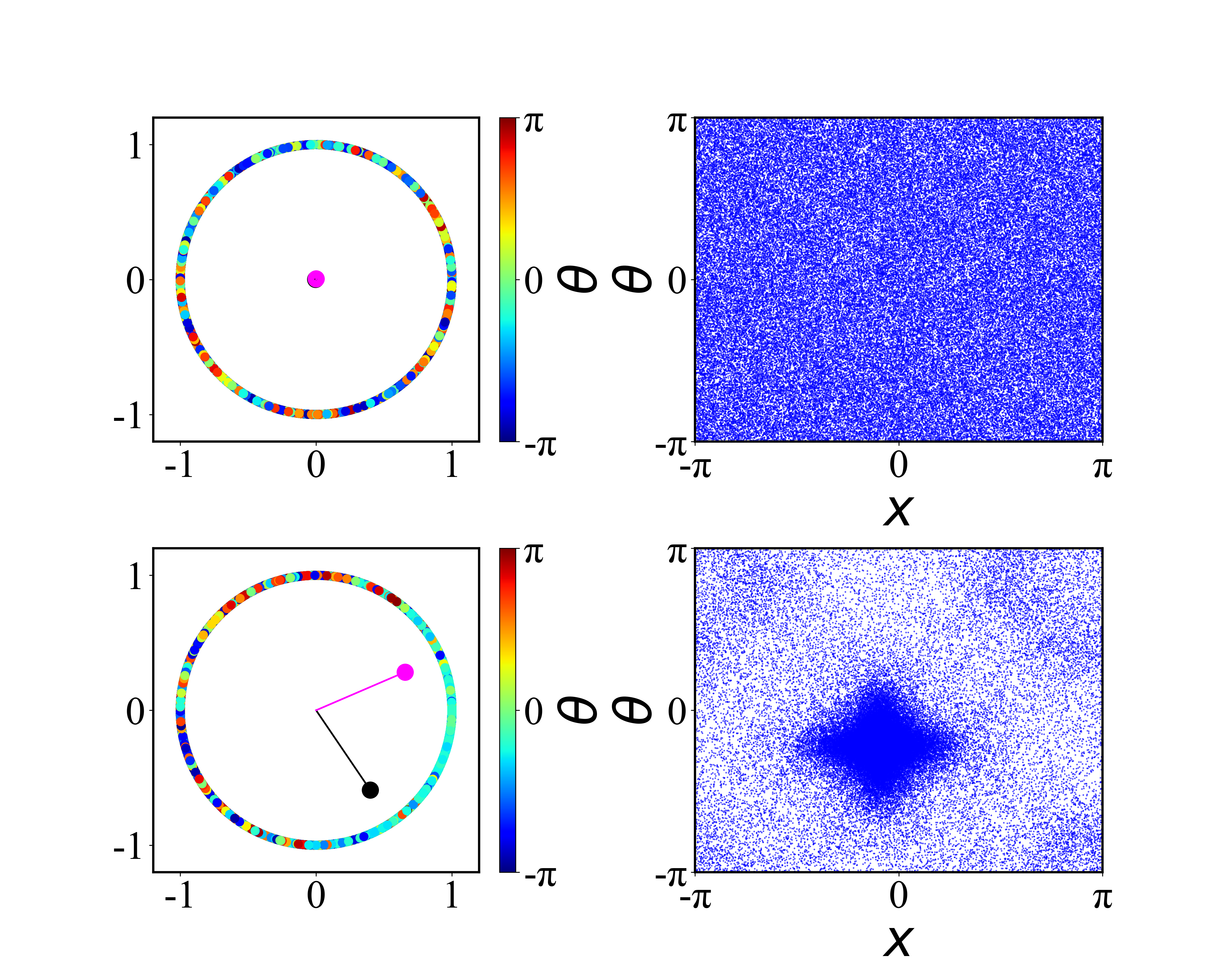}}
    \caption{{\bf Bistablilty between async and sync state}. Scatter plot for async $(0,0)$ and sync $(S, S)$ states at $K_{1}=J_{1}=3.5$, and $K_{2}=J_{2}=9$ [drawn from the region $(II)$ in Fig. \ref{j=k_fwd_bwd}(a)] is depicted in the upper and lower row, respectively. The order parameter $S_{1}^{+}(S_{1}^{-})$ is indicated by the black (magenta) circle in the left column, where the values of $S_{1}^{+}$ and $S_{1}^{-}$ are represented by the length of the line joining the center with the respective circles. Clearly, in the upper row, the values of both order parameters are almost zero, indicating the async $(0,0)$ state, while in the lower row, both the order parameters take non-zero equal value, characterizing the sync $(S, S)$ state. In the right panel, the corresponding scatter plots in the $(x,\theta)$ plane are displayed. Swarmalators are uniformly distributed in both phase and space, characterizing the async state (in the upper row), whereas in the lower row, swarmalators are locked in both phase and space and thus correspond to the sync state.}
    \label{async_sync_snapshot}
\end{figure}

In this case, the coupled complex-valued differential Eqs. \eqref{alpha_eq} and \eqref{beta_eq} become decoupled as follows
\begin{subequations}
\begin{equation} \label{j=k_alpha_eq}
    \begin{array}{l}
        \dot{\alpha} = - \mathrm{i} (v+\omega) \alpha + \dfrac{1}{2} [(H_{1}^{+})^{*}-H_{1}^{+}\alpha^{2}],
    \end{array}
\end{equation}
\begin{equation}\label{j=k_beta_eq}
    \begin{array}{l}
        \dot{\beta} = - \mathrm{i} (v-\omega) \beta + \dfrac{1}{2} [(H_{2}^{+})^{*}-H_{2}^{+}\beta^{2}],
    \end{array}
\end{equation}
\end{subequations}
and consequently the phases $\xi$ and $\eta$ develop independently analogous to typical higher-order Kuramoto model \cite{skardal2020higher}. From Eqs. \eqref{j=k_alpha_eq} and \eqref{j=k_beta_eq} one can observe that the functions $\alpha$ and $\beta$ are dependent on $(v+\omega)$ and $(v-\omega)$, respectively, which are distributed in accordance with a Lorentzian distribution characterized by a spread of $(\Delta_v +\Delta_{\omega})$. This allows us to evaluate the integrals for the order parameters $Z_{1}^{\pm}$ explicitly as
$Z_{1}^{+}=\alpha^{*}(i\Delta_{v}+i\Delta_{\omega},t)$ and $Z_{1}^{-}=\beta^{*}(i\Delta_{v}+i\Delta_{\omega},t)$ using the residue theorem. Similarly the other order parameters $Z_{m}^{\pm}$ can be obtained explicitly as $Z_{2}^{+}=\alpha^{{*}^{2}}(i\Delta_{v}+i\Delta_{\omega},t)=(Z_{1}^{+})^{2}$ and $Z_{2}^{-}=\beta^{{*}^{2}}(i\Delta_{v}+i\Delta_{\omega},t)=(Z_{1}^{+})^{2}$. Substituting these into Eqs. \eqref{j=k_alpha_eq} and \eqref{j=k_beta_eq} give 
\begin{equation}
    \begin{array}{l}
         \dot{Z_{1}}^{\pm}= -(\Delta_{v}+\Delta_{\omega})Z_{1}^{\pm} + \dfrac{1}{2} \biggl[  \biggl\{K_{1}Z_{1}^{\pm}+K_{2}(Z_{1}^{\pm})^{2}(Z_{1}^{\pm})^{*} \biggr\} 
         \\\\ ~~~~~~~~~~~~~~~~ - \biggl\{K_{1}(Z_{1}^{\pm})^{*}+K_{2}(Z_{1}^{\pm})^{{*}^{2}}(Z_{1}^{\pm})\biggr\}(Z_{1}^{\pm})^{2}  \biggr]. 
    \end{array}
\end{equation}
As anticipated, the above equations are similar to those of the typical higher-order Kuramoto model \cite{skardal2020higher} with Lorentzian natural frequency having half width $(\Delta_{v}+\Delta_{\omega})$. Assuming $\psi_{1}^{\pm}$ to be zero, we can obtain the steady state solution for the order parameters corresponding to the sync state $S_{1}^{+}=S_{1}^{-}=S$ as  
\begin{equation} \label{J=K_sync_solution}
    \begin{array}{l}
      S= {\sqrt{\frac{K_{2}-K_{1}\pm \sqrt{\left(K_{1}+K_{2}\right)^{2}-8 K_{2} \tilde{\Delta }}}{2 K_{2}}}},
    \end{array}
\end{equation}
where plus and minus signs correspond to the stable and unstable solutions if they exist. Eq. \eqref{J=K_sync_solution} resembles the solution for phase wave state, given by Eq. \eqref{phase_wave_S+} when $J_{1}=K_{1}$ and $J_{2}=K_{2}$. This suggests that the transition from async $(0,0)$ state to sync $(S, S)$ state can emerge without experiencing an intermediate phase wave $(S,0)$ or $(0, S)$ state and the forward transition occurs at the same critical coupling given by  
\begin{equation}\label{j=k_fwd_critical}
    \begin{array}{l}
          J_{1}=K_{1}=2(\Delta_{v}+\Delta_{\omega}).
    \end{array}
\end{equation}    
To illustrate this, in Fig. \ref{j=k_fwd_bwd}(a) we plot the order parameters $S_{1}^{\pm}$ as a function of $K_{1}$ with fixed higher-order coupling strengths $J_{2}=K_{2}=9$. The solid and dashed magenta curves depict the analytical predictions provided by Eq. \eqref{J=K_sync_solution} for the stable and unstable branches, respectively, which are in good agreement with the outcomes obtained through direct simulation (represented by circles). With increasing $K_{1}$, we observe a sudden transition from $S_{1}^{\pm} \approx 0$ (async state) to $S_{1}^{\pm} \approx 1$ (sync state) at $K_{1}=4$ [given by Eq. \eqref{j=k_fwd_critical}]. Another abrupt transition from $S_{1}^{\pm} \approx 1$ to $S_{1}^{\pm} \approx 0$ emerges at $K_{1}=3$ [obtained from Eq. \eqref{phase_bwd_critical} for $J_{1}=K_{1}$ and $J_{2}=K_{2}$] as $K_{1}$ is decreased adiabatically starting from the sync state. Thus, within the domain bounded by these two sudden transitions, the system exhibits a bistable nature where both the sync and async states are achievable for any specific value of coupling strength. To highlight the presence of bistability, we showcase the two distinct states for $K_{1}=3.5$ in Fig. \ref{async_sync_snapshot}: asynchronous state (upper row) and synchronous state (lower row). Besides, in Fig. \ref{j=k_fwd_bwd}(b) we plot the stability diagram for the system, which shows that beyond $K_{2}=4$, a bistable dynamics emerges in the system due to the interplay between saddle-node and subcritical pitchfork bifurcations, and the associated region of bistability $(II)$ is bounded by the curves of these bifurcations. On the other hand, for $K_{2}<4$, a continuous transition from async state $(I)$ to sync state $(III)$ occurs at $K_{1}=4$ through a supercritical pitchfork bifurcation. The stretch of the bistability region $(II)$ in the regime $K_{1}<0$ reveals the fact that for sufficiently larger higher-order coupling strengths, the system can achieve a stable sync state even when the pairwise interactions are repulsive (i.e., $K_{1}=J_{1}<0$).         

\subsubsection{General case: $J_{1}\neq K_{1}$ and $J_{2}\neq K_{2}$} 

\begin{figure*}
    \centerline{
    \includegraphics[scale=0.55]{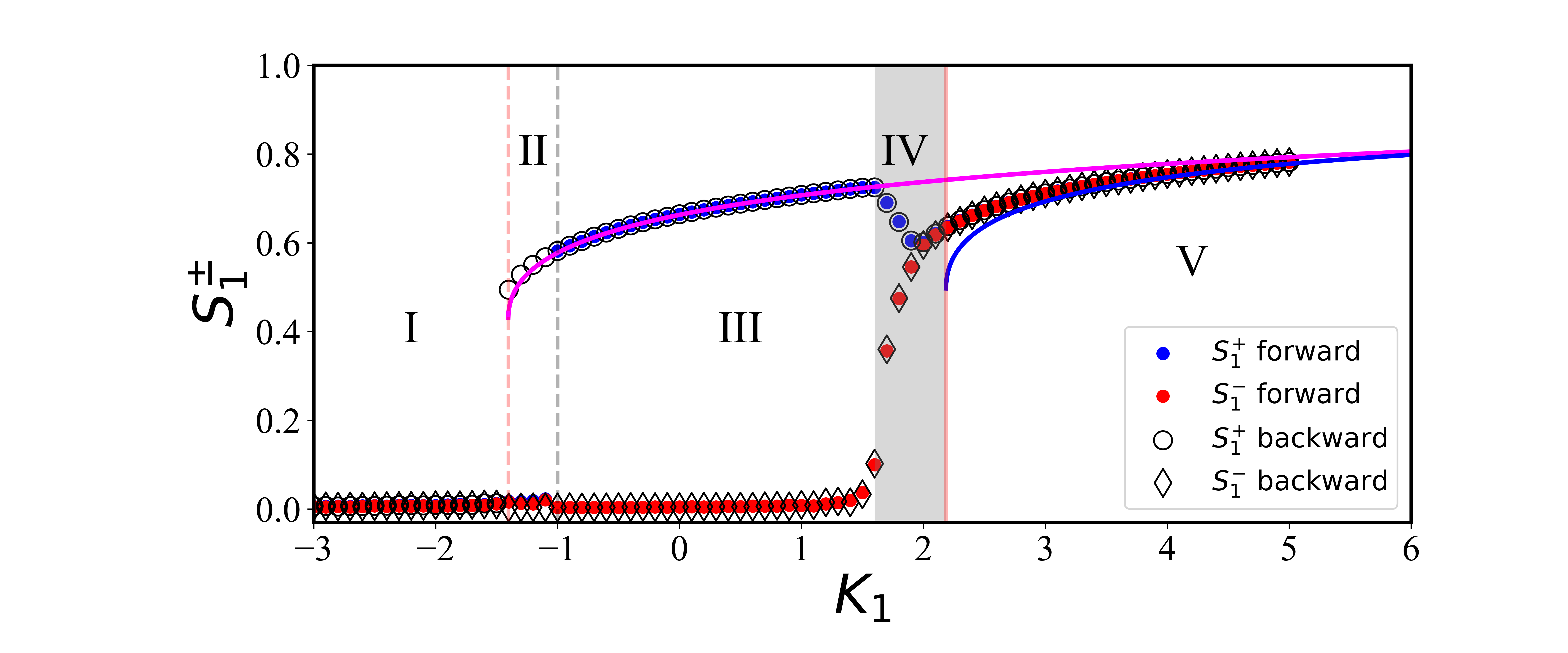}}
    \caption{{\bf Transition from async to sync state through intermediate mixed state}. Order parameter $S_{1}^{\pm}$ as a function of pairwise coupling strength $K_{1}$ for $J_{1}=9$ and fixed three-body coupling strengths $J_{2}=6.5$, $K_{2}=5.5$. Solid magenta and blue curves represent the stable solutions for phase wave and sync states, given by Eqs.\eqref{phase_wave_S+} and \eqref{general_sync_solution}, respectively. Solid circles depict the result obtained from the direct simulation of Eqs. \eqref{sw_position_eq} and \eqref{sw_phase_eq} for $N=10^{5}$ oscillators with half widths of the Lorentzian distribution $\Delta_{v}=\Delta_{\omega}=1$. An abrupt transition from the async $(I)$ to phase wave $(III)$ state is observed, which creates a domain of bistability $(II)$, where both the states are stable. On the other hand, from the phase wave state $(III)$ to sync state $(V)$, a smooth (continuous) transition occurs, resulting in the shaded region $(IV)$, which corresponds to $(S', S'')$ state, with $S'>S''$, and called as mixed state. }
    \label{gen_sync_1}
\end{figure*}

In contrast to the earlier scenario, in this case, the coupled complex-valued differential equations \eqref{alpha_eq} and \eqref{beta_eq} cannot be separated from each other. As a result, it's not possible to directly express the order parameters $Z_{1}^{\pm}$ in terms of $\alpha$ and $\beta$, respectively. Therefore, we look for a solution of Eqs. \eqref{alpha_eq} and \eqref{beta_eq} that satisfy $\dot{\alpha}\neq 0,\dot{\beta} \neq 0,Z_{m}^{\pm} \neq 0$. Assuming $\psi_{m}^{\pm}=0$, we find

\begin{subequations}\label{gen_sync_alpha_beta}
\begin{equation}\label{gen_sync_alpha}
    \begin{array}{l}
         \alpha(v,\omega)=\mathcal{F}\biggl[ \dfrac{A_{1}+A_{2}(S_{1}^{-})^{2}}{S_{1}^{+} \{C(S_{1}^{{+}^{2}}+S_{1}^{{-}^{2}})+2(D+ES_{1}^{+^{2}}S_{1}^{-^{2}})\}    }  \biggr],
    \end{array}
\end{equation}
\begin{equation}\label{gen_sync_beta}
    \begin{array}{l}
         \beta(v,\omega)=\mathcal{F}\biggl[ \dfrac{B_{1}+B_{2}(S_{1}^{+})^{2}}{S_{1}^{-} \{C(S_{1}^{{+}^{2}}+S_{1}^{{-}^{2}})+2(D+ES_{1}^{+^{2}}S_{1}^{-^{2}})\}    }  \biggr],
    \end{array}
\end{equation}  
\end{subequations}
where
\begin{equation}
    \begin{array}{l}
    A_{1,2}=2(v K_{1,2} + \omega J_{1,2}), \; B_{1,2}=2(v K_{1,2} - \omega J_{1,2}), \\ C=J_{1}K_{2}+J_{2}K_{1}, \; D=J_{1}K_{1}, \; E=J_{2}K_{2}. 
    \end{array}
\end{equation}
Now, we solve the integrals for the order parameters given by Eqs. \eqref{alpha_order_para} and \eqref{beta_order_para} using residue theorem. Further, considering the fact that in the sync state $S_{1}^{+}=S_{1}^{-}=S$, we obtain
\begin{equation}\label{gen_sync}
    \begin{array}{l}
         S=\mathcal{F}^{*}\biggl[ \dfrac{\Delta_{1}+\Delta_{2}S^{2}}{S \{2CS^{2}+2(D+ES^{4})\}} \biggr],
    \end{array}
\end{equation}
where $\Delta_{1,2}=2(\Delta_{v} K_{1,2} + \Delta_{\omega} J_{1,2})$. Eq. \eqref{gen_sync} eventually provides the expression for the order parameter, which can be acquired by solving the following equation for $S$   
\begin{equation}\label{general_sync_solution}
    \begin{array}{l}
         Ey^{3}+(C-E)y^{2}+(\Delta_{2}-C+D)y+(\Delta_{1}-D)=0, 
    \end{array}
\end{equation}
where $y=S^2$. From Eq. \eqref{general_sync_solution}, it is clear that $S$ bifurcates from zero at $\Delta_{1}=D$, i.e., 
\begin{equation}\label{gen_sync_critical}
    \begin{array}{l}
         2\biggl(\dfrac{\Delta_{v}}{J_{1}}+\dfrac{\Delta_{v}}{K_{1}}\biggr)=1. 
    \end{array}
\end{equation}
Notice that for $J_{1}=K_{1}$, the condition \eqref{gen_sync_critical} coincides with the critical coupling condition \eqref{j=k_fwd_critical} and thus once again guarantees that the sync state can bifurcate directly from the async state without passing through any intermediate state [Fig. \ref{j=k_fwd_bwd}]. However, in general circumstances (i.e., $J_{1}\neq K_{1}, J_{2}\neq K_{2}$), the transition from async state to sync state is not direct but occurs by passing through intermediate states, namely phase wave state and mixed state. To describe this, in Fig. \ref{gen_sync_1} we plot the order parameters $S_{1}^{\pm}$ for varying $K_{1}$ while keeping the other couplings fixed at nominal values $J_{1}=9, J_{2}=6.5,\;\mbox{and}\;K_{2}=5.5$. The solid magenta and blue lines correspond to the analytical predictions of stable solutions of phase and sync states obtained from Eqs. \eqref{phase_wave_S+} and \eqref{general_sync_solution}, respectively, which are in favorable alignment with the results obtained through direct simulation (shown in solid and void markers) for the finite-size system. As observed, with increasing $K_{1}$, the system passes from the async state $(I)$ to the sync state $(V)$ through the intermediate phase wave state $(III)$ and mixed state $(IV)$. Here, the evolution of the order parameters $(S_{1}^{+}, S_{1}^{-})$ shows that with increasing $K_{1}$, the system passes from $(0,0)$ (i.e., async) state to $(S,0)$ (i.e., phase wave) state through an abrupt transition, resulting into an interval of bistability $(II)$ where both the async and phase wave states are stable. On the other hand, the system goes through a continuous transition from $(S,0)$ (i.e., phase wave) state to $(S, S)$ (i.e., sync) state, resulting in the intermediate mixed state $(S', S'')$ where $S'>S''$.        

\subsection{Mixed State}\label{mixed}
In the mixed state we have $(S_{1}^{+},S_{1}^{-})=(S',S'')$, with $S'\neq S''$. This state resides as an intermediary between the phase wave and sync states and bifurcates from the $(S,0)$, or $(0, S)$ state when $S'>S''$ or $S''>S'$, through a continuous transition to the $(S, S)$ state. In Fig. \ref{gen_sync_1}, the shaded black region $(IV)$ depicts the interval of $K_{1}$ (keeping the other coupling fixed at a nominal value) for which the system passes through the mixed state. Unfortunately, we are unable to find the analytical boundaries in terms of the coupling strengths for the mixed state, and thus obtain the domain of mixed state through the numerical investigations by evaluating the order parameters $S_{1}^{\pm}$ with $S_{1}^{+}>S_{1}^{-}$. The distinctive characteristic of the mixed state lies in the fact that while both $S'$ and $S''$ remain time-independent, the functions $\alpha$ and $\beta$ exhibit time dependency \cite{yoon2022sync}. This contrasts with the time-independent equations \eqref{phase_alpha_eq} and \eqref{gen_sync_alpha_beta} associated with the phase wave and sync states, respectively.

\subsection{Abrupt transition from phase wave state to sync state}
\begin{figure*}
    \centerline{
    \includegraphics[scale=0.55]{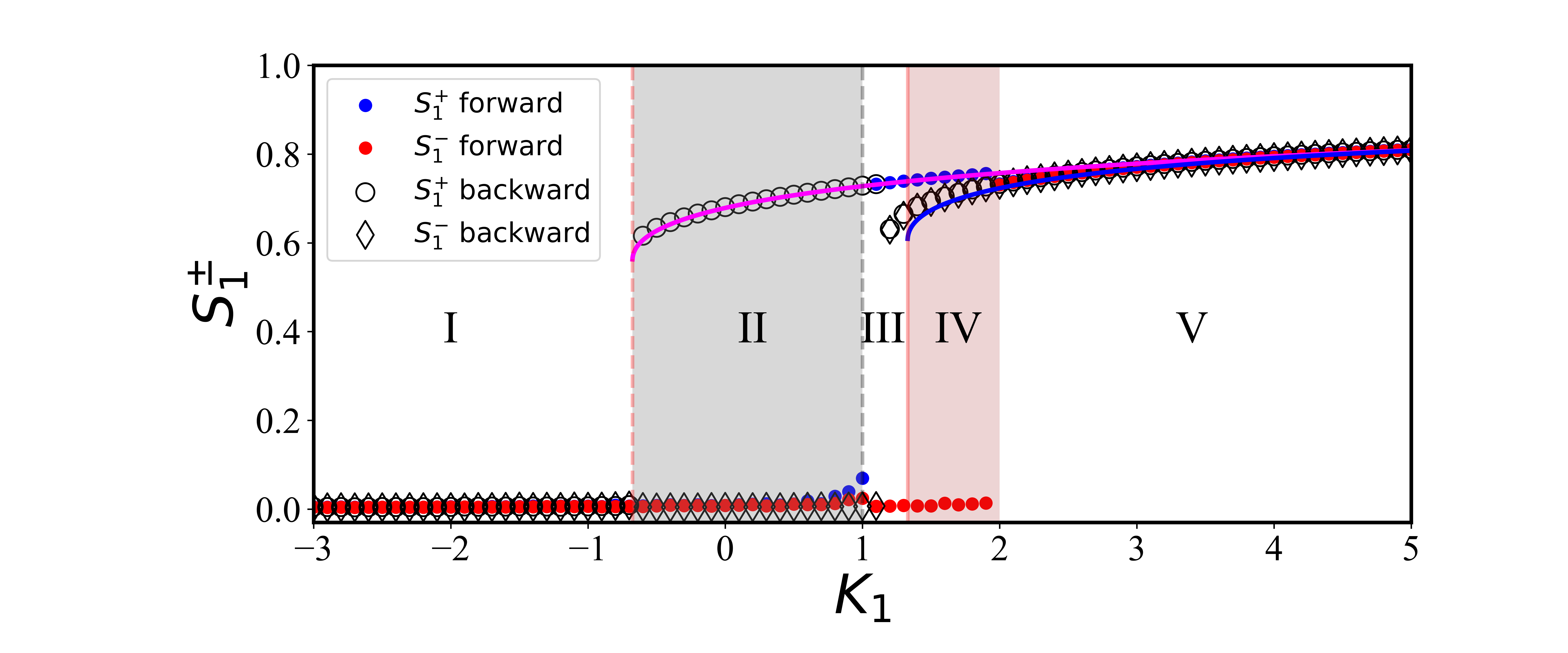}}
    \caption{{\bf Abrupt transitions from async to phase wave state, and phase wave to sync state.}. Order parameter $S_{1}^{\pm}$ as a function of pairwise coupling strength $K_{1}$ for $J_{1}=7$ and fixed three-body coupling strengths $J_{2}=8$, $K_{2}=9$. Solid magenta and blue curves represent the stable solutions for phase wave and sync states, given by Eqs.\eqref{phase_wave_S+} and \eqref{general_sync_solution}, respectively. Solid circles depict the result obtained from the direct simulation of Eqs. \eqref{sw_position_eq} and \eqref{sw_phase_eq} for $N=10^{5}$ oscillators with half widths of the Lorentzian distribution $\Delta_{v}=\Delta_{\omega}=1$. Two distinct abrupt transition is revealed. First, an abrupt transition from the async $(I)$ to phase wave $(III)$ state is observed, which creates a domain of bistability $(II)$, where both the async and phase wave states are stable. Second, from the phase wave state $(III)$ to sync state $(V)$, generating a bistable domain $(IV)$, where both the phase wave and sync states are stable.}
    \label{gen_sync_2}
\end{figure*}

\begin{figure}
    \centerline{
    \includegraphics[scale=0.26]{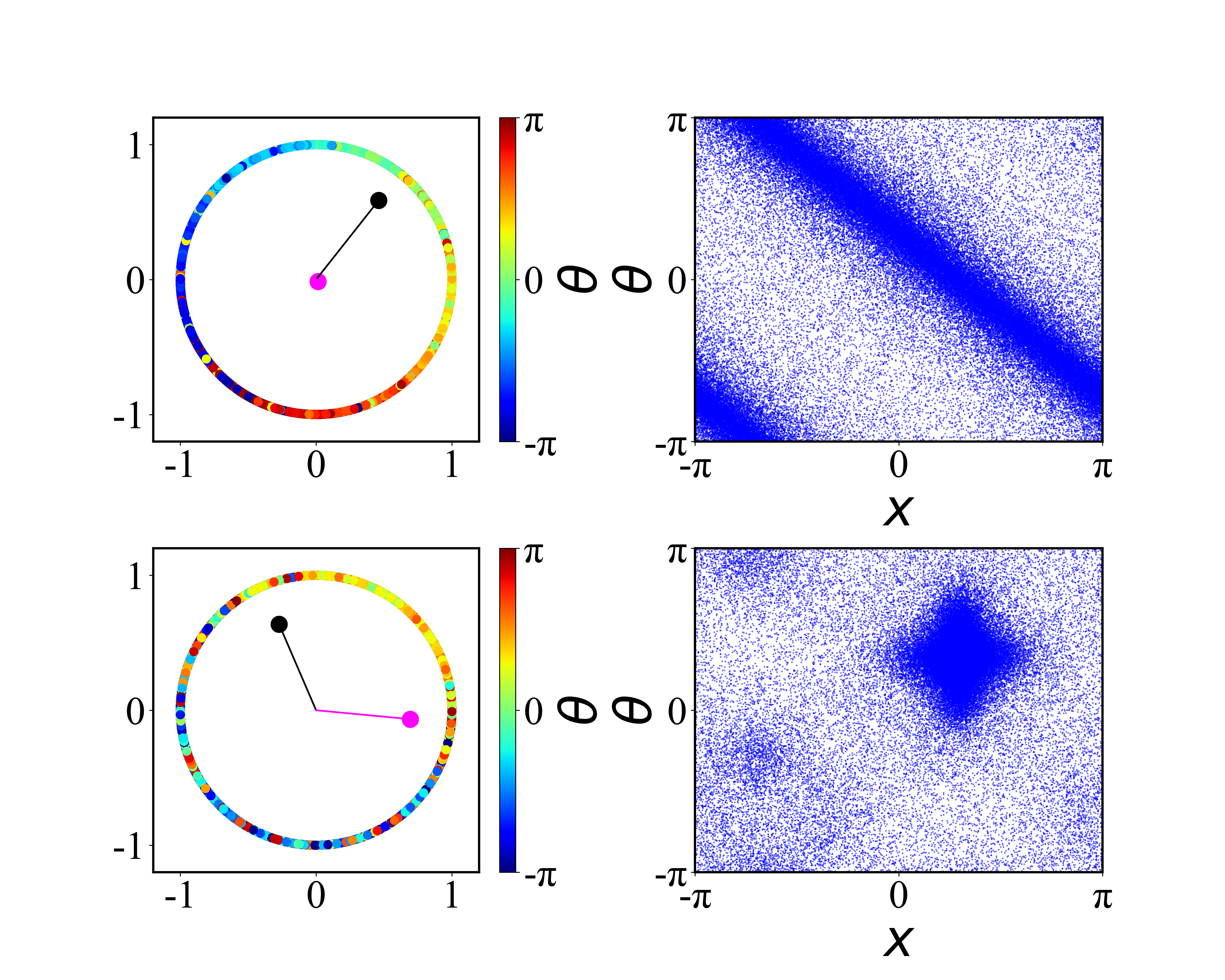}}
    \caption{{\bf Bistablilty between phase wave and sync state}. Scatter plot for phase wave $(S,0)$ and sync $(S, S)$ states at $K_{1}=1.5$, $J_{1}=3.5$, $K_{2}=9$, and $J_{2}=8$ [drawn from the region $(IV)$ in Fig. \ref{gen_sync_2}] is depicted in the upper and lower row, respectively. The order parameter $S_{1}^{+}(S_{1}^{-})$ is indicated by the black (magenta) circle in the left column, where the values of $S_{1}^{+}$ and $S_{1}^{-}$ are represented by the length of the line joining the center with the respective circles. Clearly, in the upper row, the value of $S_{1}^{+}$ is non-zero, and $S_{1}^{-}$ is zero, representing the phase wave $(S,0)$ state, while in the lower row, both the order parameters take non-zero equal value, characterizing the sync $(S, S)$ state. In the right panel, the corresponding scatter plots in the $(x,\theta)$ plane are displayed. Swarmalators display a correlation between phase and space in the upper row, characterizing the async state, whereas, in the lower row, swarmalators are locked in both phase and space and thus correspond to the sync state.}
    \label{phase_sync}
\end{figure}

As discussed above, in general, with increasing coupling strengths, a continuous transition from the phase wave state to the sync state emerges in the system through an intermediate mixed state. This smooth transition phenomenon is consistent with previously acquired results for only pairwise interactions between the swarmalators \cite{yoon2022sync}. When the higher-order coupling strengths are relatively small, a comparable transition phenomenon is still evident in our present system with higher-order interactions [Fig. \ref{gen_sync_1}]. Conversely, when sufficiently significant higher-order couplings come into play, a critical observation comes to light. In this scenario, the influence of higher-order interactions leads to an abrupt and noteworthy transition from the $(S,0)/(0, S)$ state to the $(S, S)$ state, bypassing the intermediate mixed state. To illustrate this, in Fig. \ref{gen_sync_2} we plot the order parameters $(S_{1}^{+},S_{1}^{-})$ as a function of $K_{1}$ for adequately large higher-order couplings $J_{2}=8$ and $K_{2}=9$, keeping $J_{1}$ fixed at $J_{1}=7$. The solid magenta and blue lines represent the analytically predicted stable solutions for the phase and synchronized states, respectively, derived from Eqs. \eqref{phase_wave_S+} and \eqref{general_sync_solution}. Impressively, these analytical predictions closely match the outcomes obtained through direct simulations on the finite size system, depicted by the solid and void markers. As $K_{1}$ is being first adiabatically increased to a large value and then decreased, we observe two different abrupt transitions, resulting in two distinct bistable domains. The first one corresponds to a sudden transition from the $(0,0)$ state $(I)$ to the $(S,0)$ state $(III)$, inducing a region of bistability $(II)$, where both the async and phase wave states are stable (shaded grey region). This bistable phenomenon has already been discussed in Sec. \ref{phase_wave}. The second one is associated with an abrupt transition from $(S,0)$ state $(III)$ to $(S, S)$ state $(V)$, giving rise to a bistable domain $(IV)$, where both stable phase wave and sync state can emerge (shaded red region). To elucidate this bistability nature, in Fig. \ref{phase_sync}, we demonstrate the two distinct states: phase wave state (upper row) and sync state (lower row) for fixed $K_{1}=1.5$, drawn from the region of bistability $(IV)$.        

 \section{Discussions} \label{conclusion}
 Summing up, here we have introduced an analytically tractable model of swarmalators that incorporates both pairwise and higher-order interactions (specifically three-body interactions), embedded in a simplicial complex at the microscopic level. This proposed model exhibits a high degree of complexity with four distinct collective states, namely async, phase wave, mixed, and sync states. The higher-order interactions introduce supplementary layers of nonlinearity into the behavior of the microscopic system dynamics. As a result, a few pivotal phenomena emerge that remain absent when interactions between the swarmalators are confined to only pairwise connections and lack the influence of higher-order interactions. We observe that the inclusion of higher-order interactions leads to abrupt transitions from the async state to either the phase wave state or the sync state, depending on the specific coupling strength configurations. Our observations also reveal that when the higher-order interactions are sufficiently strong, phase wave and sync state can emerge and persist, even in scenarios where pairwise couplings are repulsive. This implies that despite the potential decay of specific coupling types, the existence of alternative forms of coupling can play a pivotal role in upholding the regimes of bistability between async and phase wave (sync) states. Furthermore, our findings extend to the discovery that substantial higher-order couplings can facilitate a direct emergence of the sync state from the phase wave state without passing through the mixed state. This distinct behavior stands in contrast to situations involving solely pairwise interactions, where the synchronized state bifurcates from the phase wave state through the intermediate mixed state. 
 \par Thus, our study makes a substantial contribution to understanding the influence of higher-order interactions on shaping the collective dynamics of swarmalators, although numerous avenues for further exploration remain open. In this context, we specifically focus on higher-order interactions up to the third order, employing an all-to-all coupling arrangement. Consequently, investigating interactions beyond the three-body and exploring localized coupling configurations becomes a particularly intriguing prospect for future research. It is also worth noting that our current investigation is limited to analyzing the effect of higher-order interactions solely within a 1D swarmalator model on a ring. As such, an equally captivating avenue for future exploration lies in examining how higher-order interactions impact the collective behavior within 2D and other higher-dimensional swarmalator systems. 
	
\section*{Acknowledgements}
M.S.A. and G.K.S. would like to thank Kevin O'Keeffe for his valuable comments. M.P. is supported by the Slovenian Research Agency (Grant P1-0403). 


     
\bibliographystyle{apsrev4-1} 
\bibliography{hoi_swarm}
	
\end{document}